\documentclass[10pt, 3p]{elsarticle}

\usepackage[british]{babel}
\usepackage[utf8]{inputenc}
\usepackage{graphics}
\usepackage{graphicx}
\usepackage{amsmath,amssymb,esint}
\usepackage{lineno}
\usepackage{multirow}
\usepackage{subfig}
\usepackage{natbib}
\usepackage{eucal}
\usepackage{mathtools}
\usepackage{placeins}
\usepackage{enumerate}
\usepackage[usenames,dvipsnames]{xcolor}
\usepackage[colorlinks]{hyperref}
\usepackage{setspace}
\usepackage{MnSymbol}
\usepackage[linesnumbered,ruled,vlined]{algorithm2e}
\usepackage{array}

\biboptions{sort&compress}
\graphicspath{{./}{figures/}}

\journal{}
\makeatletter
\def\ps@pprintTitle{%
  \let\@oddhead\@empty
  \let\@evenhead\@empty
  \let\@oddfoot\@empty
  \let\@evenfoot\@oddfoot
}
\makeatother

\begin{document}

\begin{frontmatter}

\title{A marching cubes based method for topology changes in three-dimensional two-phase flows with front tracking}

\author[affil1]{Gabriele Gennari}
\author[affil1]{Christian Gorges}
\author[affil2]{Fabian Denner}
\author[affil1]{Berend van Wachem\corref{cor1}}
\ead{berend.van.wachem@multiflow.org}

\address[affil1]{Chair of Mechanical Process Engineering,
Otto-von-Guericke-Universit\"{a}t Magdeburg,\\ Universit\"atsplatz 2, 39106 Magdeburg, Germany}
\address[affil2]{Department of Mechanical Engineering, Polytechnique Montr\'eal,\\ Montr\'eal, H3T 1J4, QC, Canada}

\cortext[cor1]{Corresponding author: }

\begin{abstract}
The handling of topology changes in two-phase flows, such as breakup or coalescence of interfaces, with front tracking is a well-known problem that requires an additional effort to perform explicit manipulations of the Lagrangian front. In this work, we present an approach that allows to perform topology changes with interfaces made of connected triangular elements. The methodology consists of replacing the fluid entities that undergo breakup/coalescence with the iso-surface corresponding to the indicator function value $\mathcal{I} = 0.5$, which automatically returns the shape of the bodies after topology changes. The generation and triangulation of such surface is obtained by exploiting the marching cubes algorithm. Since we perform the reconstruction of the interface only for the bodies that experience breakup/coalescence, the increase in computational cost with respect to a classic front tracking scheme without topology changes is small. Using validation cases, we show that the proposed reconstruction procedure is second-order accurate for volume conservation and able to capture the physics of several two-phase flow configurations undergoing topology changes. The validation cases include the breakup of a droplet in simple shear flow and two rising bubbles in different regimes (peripheral and central breakups). Coalescence is tested by modelling the binary collision between two droplets. For the selected validation cases, an excellent agreement between the numerical results and experiments is observed. The proposed methodology is able to capture the details of such interfacial flows, by predicting accurately the coalescence/breakup dynamics, as well as the number, size and shapes of satellite droplets/bubbles after topology changes.
\end{abstract}
\end{frontmatter}

\section{Introduction}
The numerical modelling of interfacial flows poses several challenges due to complex phenomena that occur at the interface between two different fluids. Such difficulties include the accurate representation of the interface and its transport within the domain, the computation of interfacial effects (e.g., surface tension force, mass transfer) and the changes in the topology of the interface due to breakup or coalescence of fluid entities (e.g., droplets, bubbles) or ligaments (e.g., atomization of liquid jets). Several methodologies have been developed to address the above difficulties, and can be classified into front capturing and front tracking techniques. 

The first class of methods relies on the transport of an Eulerian field, such as the colour field for the volume of fluid (VOF) approach \citep{Hirt1981, Scardovelli1999} and the distance function in the level set (LS) method \cite{Sussman1994}, and, in both cases, the interface is implicitly captured by transporting a corresponding field on the Eulerian mesh. The VOF method exhibits excellent mass conservation properties, but the computation of the interface curvature is not always straightforward, leading to the well-known effect of spurious currents \citep{Harvie2006}. The LS method offers greater accuracy in computing the curvature but does not perform as well as the VOF in terms of mass conservation. The second class of methodologies, i.e., front tracking, uses an auxiliary surface mesh to describe the interface \citep{Unverdi1992, Popinet1999, Glimm1998, Tryggvason2001}. Such mesh consists of a triangular (in 3D) front, where each vertex (also referred to as marker) is advected in a Lagrangian fashion with the underlying velocity field computed on the Eulerian grid. Front tracking usually comes with additional memory requirements to store the connectivity of the front (i.e., the connections between vertices, edges and triangles of the Lagrangian mesh) and computational cost for the remeshing of the front. Since the markers are advected with the flow velocity field, they can accumulate or their density can decrease in specific regions of the front (e.g., bottom and top sides of a rising bubble, respectively). Such events can lead to wrinkles in the surface (accumulations of markers) or to under-resolved interfaces in the opposite scenario. Remeshing is necessary to avoid these phenomena and ensure a good quality of the Lagrangian mesh. However, remeshing can introduce shape and volume errors, and specific vertex repositioning algorithms for front tracking have been developed to mitigate these effects \citep{Gorges2022}. Despite the additional complexity of front tracking, this methodology is widely used for the modelling of two-phase flows. The main strengths of front tracking come from the explicit representation of the interface, which allows to reach higher accuracy in the computation of the properties of the front (e.g., normal vector and curvature) and greatly simplifies the treatment of surface scalar fields, such as the concentration of insoluble surface-active substances \citep{Muradoglu2008, Muradoglu2014}. 

Research on both classes of numerical methods has mainly focused on the development of schemes for the advection of the interface or the computation of surface tension effects \citep{Popinet2018}, whilst advances in the modelling of topology changes has lagged far behind. Front capturing and front tracking schemes behave in a very different way with regards to topology changes. In the VOF or LS methods without special treatments, coalescence or breakup occur numerically whenever the distance between two interfaces becomes smaller than the grid size. This happens because the Eulerian field that implicitly represents the interface cannot distinguish multiple interfaces within the same computational cell. Although this comes with the advantage that no additional effort is required to obtain a change in the topology of the interface, the eventual breakup/coalescence is a numerical artefact, i.e., it depends solely on the grid resolution and does not take into account the physics of the problem. Ad hoc techniques have been proposed to handle topology changes for specific configurations with front capturing methods. A simple approach consists of using multiple colour functions for each fluid particle. In this way, when two different interfaces come closer than one cell, they do not merge because each interface is represented by a different field \citep{Coyajee2009, Balcazar2015}. To reduce the number of fields necessary to track all the fluid particles (which can be prohibitive when a large number of bubbles/droplets is simulated), a multilayer VOF method is presented by \citet{Karnakov2022a}. In the case of thin ligaments that undergo breakup, a novel algorithm is proposed by \citet{Chirco2022}, that allows to detect thin structures and perforate them before their thickness reaches the grid size threshold. Recently, extension to classic VOF schemes have been proposed that allow to describe two planar interfaces in the same computational cells, paving the way towards the modelling of sub-grid films \citep{Han2024}.

Front tracking methods behave in the opposite way, i.e., coalescence or breakup of interfaces do not occur spontaneously, regardless of the distance between the fronts and the resolution of the Eulerian mesh. Therefore, topology changes must be taken into account explicitly by performing a reconstruction of the Lagrangian front. The methodologies available can be classified into grid-based or grid-free approaches, depending on whether they rely on an auxiliary (Eulerian) mesh \citep{Regnault2024}. Grid-free approaches consist of manipulating the front elements close to the region where breakup or coalescence occur, which, therefore requires an algorithm to identify such areas. A simple implementation for droplets collision is presented by \citet{Nobari1996a} and \citet{Nobari1996} for 2D and 3D simulations, respectively. Their approach consists of removing triangular elements from both fronts when they are close to each other and nearly parallel. The holes generated in both droplets are then reconnected in order to form a single surface. Resolutions of intersections between fronts that lead to changes in the front topology are achieved by \citet{Glimm1998a} with an untangle algorithm in a grid-free framework. This algorithm consists of finding the cross points (in 2D) of the two fronts and removing the non-physical regions between the curves. Extension to 3D is presented in \citet{Glimm2000}. 

More recent advancements in grid-free approaches are presented by \citet{Razizadeh2018} and \citet{Regnault2024}. The former presents a methodology that consists of identifying pairs of triangular elements with normal vectors pointing towards opposite directions and distance below a threshold value. Such elements are then collapsed by replacing each pairs of vertices with their mid point. In this way, a region of dual elements is created and, after their removal, the change in the front topology is achieved. The methodology has been successfully tested for breakup and coalescence of droplets. A similar approach is developed by \citet{Regnault2024}, where two different algorithms for breakup and coalescence are presented. Two coalescence regions on the fronts that undergo merging are obtained through projection of the triangle vertices that are closer than a threshold distance. The polygonal regions are then adapted on both interfaces so that they can be merged halfway between the fronts. The algorithm is tested for breakup and coalescence, however the author reports that the hypotheses behind the model limit its application to simple topological changes. 

Grid-based methods employ an auxiliary grid to perform changes in the topology of the interface. \citet{Glimm2000} propose a methodology based on three steps to resolve overlapping interfaces: computation of the intersection between the fronts and the auxiliary grid edges, removal of invalid intersections and reconstruction of the front into a topologically valid interface. Reconstruction is performed adopting a marching cubes scheme \citep{Lorensen1987}, i.e., based on the phase indicator function values stored at the grid vertices. The auxiliary grid is the same as the one used to solve the governing equations. \citet{Glimm2000} also combine their grid-free and grid-based methods into an hybrid approach that shows better performance. In a similar way, \citet{Du2006} and \citet{Bo2011} propose a locally grid-based method, where the front is mainly handled in a grid-free way, and grid-based reconstruction is limited to regions of the front where triangles form a topologically inconsistent surface. The main advantage of such hybrid approaches consists of combining the best features of both methods. The grid-free approach provides high-quality reconstructed fronts but it is not robust enough to handle complex topological cases. On the other hand, grid-based approaches always return a valid interface but are prone to interpolation errors and result in an excessively smoothed interface. 

Front tracking methods without connectivity, i.e., the logical connections between the vertices, edges and triangles of the Lagrangian front, have been introduced to reduce the memory requirements necessary to store the connectivity information and ease the complexity of remeshing algorithms \citep{Torres2000, Shin2002}. Such methods have also been employed to model topology changes in two-phase flows. \citet{Torres2000} develop the point-set method, a front tracking algorithm that uses unconnected interfacial markers to represent the interface. The authors derive a procedure to compute surface properties (e.g., normal vector and area) and spread the indicator function from the Lagrangian front to the Eulerian mesh via an iterative solver that comes with an additional cost compared to classic front tracking schemes. To maintain a uniform distribution of the markers, a regeneration of the interface is periodically executed. Such regeneration consists of projecting the cell centres of a Cartesian mesh onto the interface and retaining only those projected points that lie within the same cell as the centre they originate from. Such procedure allows to model coalescence or breakup by projecting points from cells outside or inside the interface respectively. In the generic case of a front undergoing both types of topology changes, a more complex regeneration procedure is required. 

Another connectivity-free technique that combines front tracking and level set methods to handle topology changes is the level contour reconstruction method (LCRM) \citep{Shin2002, Shin2005}. This approach uses a Cartesian grid to perform periodic (e.g., every 100 time steps) reconstruction of the entire Lagrangian front. Similar to the approach of \citet{Torres2000}, such reconstruction is needed to maintain a good quality of the surface mesh. The indicator function $(\mathcal{I})$ is computed at the vertices of the auxiliary mesh and the intersections between the new front and the grid edges are computed through linear interpolation from the vertex values and located where $\mathcal{I} = 0.5$. When all the intersection points are found, the old front is discarded and entirely replaced by the new one. 

Several improvements to the LCRM method have been proposed in the literature. A high-order interpolation kernel is adopted in \citet{Shin2007a} to improve the redistribution of mass in the reconstructed front that occurs between regions with different curvatures. A methodology that relies on the LCRM method only to deal with topology changes is presented by \citet{Singh2007}. Their method uses a classic front connectivity to perform remeshing operations, such as edge splitting and collapsing. A hybrid approach that combines LCRM and LS methods is proposed by \citet{Shin2009}, where a LS-like distance function $(\phi)$ is explicitly computed on the Eulerian grid from the vector distance between each front elements and the neighbouring cell centres. This enhances the accuracy of the curvature and, therefore, reduces spurious currents effects. The distance field can be used instead of the indicator function to reconstruct the front by locating the intersections at $\phi = 0$. The original LCRM method suffers from ambiguities in the reconstruction procedure whenever each edge of a cell is cut by the reconstructed interface (in 3D this applies to a face of the auxiliary grid). A tetra-marching procedure is adopted in \citet{Yoon2010} to resolve such ambiguities and consists of splitting each face of the grid used for reconstruction along one of the diagonals. The reconstruction on a triangular face is non-ambiguous, however the result in terms of interface topology depends on the direction of the split, i.e., on which of the two diagonals is used. The consequence of this is that breakup/coalescence depend on the direction used for the splitting. To minimise this effect, the direction of the diagonals can be swapped periodically during the simulation. Using a zig-zag pattern of such diagonals ensures continuity across cell boundaries of the reconstructed front. A significant improvement to the LCRM algorithm is the local front reconstruction method (LFRM) \citep{Shin2011}. This approach does not rely on an Eulerian field, but reconstructs instead directly from the Lagrangian front. The intersection points between the front and the auxiliary grid edges are computed geometrically from the front and the reconstruction operation is performed within each cell in a way that ensures local mass conservation. The handling of topology changes is equivalent to the LCRM method and suffers as well of the same non-uniqueness feature, depending on the direction of the diagonal cut. Both methodologies (LCRM, LFRM) can naturally handle coalescence/breakup of interfaces and return a result similar to front capturing methods (e.g., VOF, LS). Recent advancements in connectivity-free approaches include the edge-based interface-tracking (EBIT) method \citep{Chirco2023, Pan2024}, where the Lagrangian front consists of markers that are allowed to move exclusively along the edges of the Eulerian grid. Markers are advected with a split method and topology changes occur when two markers reach the same cell edge aligned to the advection component. Such points are then removed and the front is reconnected accordingly. The drawback of this approach is that breakup/coalescence do not occur when the fronts approach each other with a direction not aligned to the advection component.

The problem of dealing with triangular surface meshes that represent interfaces undergoing topology changes is not exclusive to computational fluid dynamics simulations. Several algorithms were also developed by researchers in computer graphics to render realistic fluid animations, and front tracking techniques have proven effective in this field. Both grid-based \citep{Muller2009a, Wojtan2009, Wojtan2010, Yu2012a} and grid-free \citep{Brochu2009, Brochu2010, Chentanez2015} approaches have been successful in computer graphics simulations. Grid-based methods often rely on the marching cubes algorithm to reconstruct the iso-surface corresponding to the interface, whereas specific approaches are devised to preserve sub-grid features \citep{Muller2009a}, confine the mesh modifications to the parts affected by breakup/coalescence \citep{Wojtan2009} and reduce errors between the Lagrangian mesh and the corresponding iso-surface \citep{Yu2012a}. An alternative to the marching cubes approach is presented by \citet{Wojtan2010}, where the interface to be reconstructed is replaced by the corresponding convex-hull shape, limiting the loss of sub-grid details. The criteria that drive the design of computer graphics applications are based on computational efficiency, {ease of parallelisation} and preservation of thin features for realistic renderings. However, an accurate representation of the interfacial phenomena is not a priority and validation of these algorithms against real cases (e.g., experiments) is missing in the literature.  

From the literature review presented above, it emerges that several non-trivial issues still need to be addressed when dealing with the modelling of topology changes with front tracking methods. In this work, we present a grid-based numerical framework based on the marching cubes algorithm for the reconstruction of the front, where the interface after breakup/coalescence is obtained by a reconstruction of the iso-surface corresponding to the indicator function value $\mathcal{I} = 0.5$. {Contrary to the LCRM/LFRM methods, where the topology of the interface is not determined a-priori, the marching cubes algorithm is based on a list of 256 pre-computed topological configurations, which limits the amount of reconstruction operations at run time and makes the algorithm efficient.} Furthermore, we retain the front connectivity between the Lagrangian elements that allows us to limit the reconstruction of the front only to instances of topology changes, whilst volume-conserving remeshing operations \citep{Gorges2022} are performed to maintain a good quality of the triangles. Each Lagrangian body is unambiguously identified by an integer number and a breakup/coalescence algorithm is introduced to quickly identify the fronts that undergo topology changes based on a distance threshold. In this way, only the selected bodies for breakup/coalescence are reconstructed, whilst the rest of the front remains unaltered. The main goal of our approach is to provide a framework to handle complex topology changes {and, at the same time, reduce volume and shape errors that are inevitably introduced when the front is entirely reconstructed. It is finally noted that a further benefit of limiting front reconstructions to the necessary minimum consists of avoiding excessive smoothing of surface scalar fields (e.g., surfactants), which require a two-way (front-to-grid and grid-to-front) communication whenever the interface is reconstructed \citep{Shin2018}.}

The following sections are organised as follows. The one-fluid formulation of the Navier-Stokes equations and the main ingredients of front tracking are presented in section \ref{sec:Governing equations and numerical framework}, whilst the core of the reconstruction procedure (i.e., marching cubes and the detection algorithm for breakup/coalescence) is presented in section \ref{sec:Front reconstruction due to topology changes}. Finally, several test cases and validation against previous experimental works are discussed in section \ref{sec:validation}.

\section{Governing equations and numerical framework}
\label{sec:Governing equations and numerical framework}
In this work we perform direct numerical simulations of incompressible two-phase flows. The system of governing equations is given by the Navier-Stokes equations, which are solved here in the corresponding one-fluid formulation:
\begin{equation}
\label{eq:continuity}
    \nabla \cdot \mathbf{u} = 0
\end{equation}
\begin{equation}
\label{eq:momentum}
    \rho \left( \frac{\partial \mathbf{u}}{\partial t} + \nabla \cdot ( \mathbf{u} \otimes \mathbf{u}) \right) = - \nabla p + \nabla \cdot (2\mu\mathbf{D}) + \rho \mathbf{g} + \mathbf{f}_{\sigma} \delta_S
\end{equation}
where $\mathbf{u}$ is the fluid velocity, $\rho$ and $\mu$ are the phase density and viscosity, respectively, $\mathbf{D}$ is the deformation tensor $\mathbf{D} = \left[ \nabla\mathbf{u} + (\nabla\mathbf{u})^T \right]/2$, $p$ is the pressure field, $\mathbf{g}$ is the gravitational acceleration and $\mathbf{f}_{\sigma}$ is the surface tension force. This term is converted into a volumetric source term via the interfacial delta function $\delta_S$, which is the surface-equivalent of the Dirac $\delta-$function concentrated on a point. Equation \eqref{eq:continuity} represents the continuity equation, whilst the conservation of momentum is given by equation \eqref{eq:momentum}. The system of governing equations is solved in a finite-volume framework with a coupled and pressure-based solver, where the primary variables are stored in a collocated arrangement \cite{Denner2014a, Denner2020}. The terms Eulerian and fixed grid will be used interchangeably in the rest of the work to refer to the mesh used for the discretization of equations (\ref{eq:continuity}, \ref{eq:momentum}). Although the methodology presented in this work can be used with any kind of Eulerian grid (i.e., structured and unstructured), only Cartesian meshes are considered here. The accuracy of the numerical schemes is second-order in space and time for the discretisation of equations \eqref{eq:continuity} and \eqref{eq:momentum} \citep{Denner2020}.

Front tracking algorithms use a Lagrangian description of the interface that consists of a triangular mesh. The position of each vertex, $\mathbf{x}_i$, is updated at every time step by solving the following transport equation:
\begin{equation}
\label{eq:front_advection}
    \frac{\mathrm{d}\mathbf{x}_i}{\mathrm{d}t} = \mathbf{\bar{u}}(\mathbf{x}_i)
\end{equation}
where the advection velocity, $\mathbf{\bar{u}}$, represents the interpolation of the velocity field, $\mathbf{u}$, from the Eulerian grid to the position of the $i$-th Lagrangian vertex. The divergence-preserving interpolation scheme \citep{Toth2002} is adopted in the present work on the basis of the results presented by \citet{Gorges2022}, who shows a significant reduction in volume errors compared to other approaches commonly used in front tracking simulations. The velocity is first computed on the Eulerian cell faces (from the cell centres) and then interpolated to the Lagrangian vertices. Correction terms are applied to the interpolation scheme in order to preserve exactly the discretised value of the divergence of the velocity field. 

The one-fluid formulation consists of a single set of governing equations that is valid in the entire domain. The scalar fields representing the fluid properties (e.g., density, viscosity) are generally discontinuous across the interface, i.e., they assume different values in each phase, and are computed through a volume-based weighted average. For density, this reads:
\begin{equation}
\label{eq:density}
    \rho(\mathbf{x}) = \rho_\mathrm{a} + \mathcal{I}(\mathbf{x}) (\rho_\mathrm{b}-\rho_\mathrm{a})
\end{equation}
and an analogous equation is used for viscosity. The variable $\mathcal{I}$ is the indicator function that identifies the regions $\left( \Omega_a, \Omega_b \right)$ of phases $\left( a, b \right)$ in the two-phase system under consideration:
\begin{equation}
\label{eq:indicator}
  \mathcal{I}(\mathbf{x}) = \begin{cases}
    0, &\mbox{if} \ \mathbf{x} \in \Omega_\mathrm{a}\\
    1, &\mbox{if} \ \mathbf{x} \in \Omega_\mathrm{b}
  \end{cases}
\end{equation} 
Contrary to the VOF method, the indicator function is not directly advected on the Eulerian grid. Instead, it is obtained by spreading the indicator gradients from the Lagrangian front. This procedure is accomplished by solving the following Poisson equation on the fixed grid \cite{Tryggvason2011a}:
\begin{equation}
\label{eq:poisson}
    \nabla^2 \mathcal{I}(\mathbf{x},t) = \nabla \cdot \left( \nabla \mathcal{I} \right) \approx \nabla \cdot \left( \sum\limits_{E} \Delta \mathcal{I} \mathbf{n}_E D(\mathbf{x} - \mathbf{x}_E) \frac{A_E}{V} \right)
\end{equation}
where the summation term on the RHS represents the (discretized) contribution of each triangle $E$ (with area $A_E$, normal vector $\mathbf{n}_E$ {and centre coordinates $\mathbf{x}_E$}) for the Eulerian cell with centre coordinates $\mathbf{x}$ and volume $V$. The jump value of the indicator function between the phases is $\left( \nabla \mathcal{I} \right) = -1$, according to equation \eqref{eq:indicator}. The distribution function $D(\mathbf{x} - \mathbf{x}_E)$ ensures that the contribution of each triangle $E$ is spread only to the neighbouring cells. This is achieved in the present work by adopting the Peskin weighting approach \cite{Peskin2003,Peskin1972} in a $5 \times 5 \times 5$ stencil centred on the Eulerian cell that contains the centre of the generic front triangle:
\begin{equation}
    d(r) = \begin{cases}
    \frac{1}{4} \left( 1 + \cos \left( \frac{1}{2} \pi r \right) \right), & |r| < 2 \\
    0, & |r| \geq 2 \\
    \end{cases}
    \label{eq:PeskinEquation}
\end{equation}

The weighted contribution of the triangle is then computed by multiplying the one-dimensional weights from equation \eqref{eq:PeskinEquation}, i.e., $D(\mathbf{x} - \mathbf{x}_E) = d(r_x) d(r_y) d(r_z)$, where $r_x$ is the non-dimensional distance, i.e., $r_x = (x - x_E) / \Delta x$. When the contributions from all the elements of the Lagrangian front are computed, the gradient of the indicator function at the generic Eulerian location $\mathbf{x}$ can be computed according to the RHS of equation \eqref{eq:poisson}, which is then integrated with a standard second-order centred finite difference scheme. A similar treatment is adopted for the surface tension force, where the corresponding term on the RHS of the momentum equation is spread from the Lagrangian front to the Eulerian mesh \citep{Tryggvason2011a}:
\begin{equation}
\label{eq:surface_tension}
    \mathbf{f}_\sigma = \sum\limits_{E} \mathbf{f}_{\sigma, E} D(\mathbf{x} - \mathbf{x}_E)
\end{equation}
where $\mathbf{f}_{\sigma, E}$ is the surface tension force on the generic front triangle $E$. Such term is computed with a line integral over the triangle perimeter $e$ as $\mathbf{f}_{\sigma, E} = \sigma \sum\limits_{e} \mathbf{p}_e l_e$. The planar vector $\mathbf{p}_e$ is obtained by the cross product between the front normal vector at edge $e$ and the tangential vector at the same edge. The length of the generic edge $e$ is $l_e$.

\section{Front reconstruction due to topology changes}
\label{sec:Front reconstruction due to topology changes}
In front tracking simulations, topology changes (e.g, breakup of bubbles, coalescence of colliding droplets) must be addressed explicitly by performing geometrical operations on the Lagrangian front. If no action is undertaken, the connectivity between the front elements remains unchanged and no modifications in the front topology are possible. 

\begin{figure}[ht]
    \centering
    \includegraphics[]{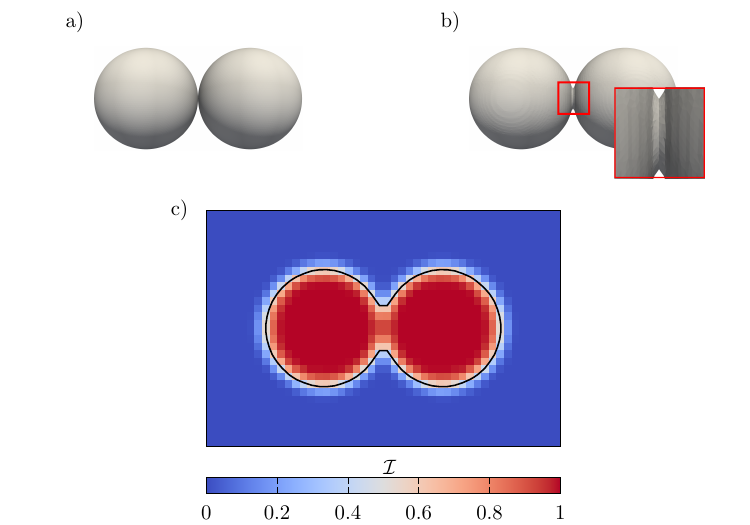}
    \caption{Overview of the reconstruction procedure: (a) original configuration, (b) reconstructed body after coalescence and (c) contours of the indicator function with the iso-surface $S_{\mathcal{I} = 0.5}$ (black line).}
    \label{fig:reconstruction_overview}
\end{figure}

In this work, a front reconstruction approach based on the indicator function (stored on the Eulerian grid) is proposed and the main idea is illustrated in Figure \ref{fig:reconstruction_overview}. 
We consider the case of two adjacent spheres that undergo coalescence. Figure \ref{fig:reconstruction_overview}a shows the initial configuration, where the contact region between the particles is (ideally) limited to one point. The corresponding indicator scalar field is shown in Figure \ref{fig:reconstruction_overview}c, where phase $\Omega_a$ (equation \eqref{eq:indicator}) is assumed to be the disperse phase (i.e., the two fluid particles). The topology of the interface can be reconstructed based on the information provided by the indicator field. Since $\mathcal{I}$ is bounded between $0 < \mathcal{I} < 1$, the interface consists of the points $\mathbf{x}$ that belong to the iso-surface $S_{\mathcal{I} = 0.5}$. For the example under consideration, a slice of the iso-surface is represented by the black line in Figure \ref{fig:reconstruction_overview}c. As can be immediately seen, the indicator-based reconstructed front automatically provides a single merged front. This result is not surprising, as it is the same mechanisms that occurs in VOF simulations, where the indicator function is explicitly advected and topology changes occur numerically whenever the distance between the fronts is approximately equal to the grid spacing. For the same reason, the reconstructed front results in a single body only if the distance between the initial spheres is smaller than grid size where $\mathcal{I}$ is stored, {otherwise the two original spheres would be recovered}. {Despite the similarities with the front capturing methods mentioned above,} it is important to point out that the front tracking approach can provide a better control over the topology change process, since the decision to trigger the reconstruction is based on user-defined criteria. It is possible, for instance, to reconstruct the front when the distance between two interfaces is arbitrary smaller than the grid size (which is generally not possible in standard front capturing methods, since a single cell cannot handle more than one interface), or introduce a physically-consistent model to predict the coalescence/breakup time (e.g., film rupture model for the coalescence of a binary droplet collision \cite{Rajkotwala2018}). Another remarkable advantage of front tracking consists of the separate (Lagrangian) mesh used for the description of the front, which allows to obtain the indicator function using equation \eqref{eq:indicator} on an auxiliary grid (different from the one used to solve the Navier-Stokes equations). In this way it is possible to increase the accuracy of the reconstructed front without affecting the computational cost of the solution of the governing equations. This feature is particularly relevant when it is essential to capture regions with strong curvatures and will be properly discussed later in section \ref{sec:Marching cubes algorithm}. 

The last step of the reconstruction procedure consists of recreating a fully-connected Lagrangian front from the iso-surface $S_{\mathcal{I} = 0.5}$. This requires the computation of the coordinates of the new vertices and the corresponding connectivity between edges and triangles. This is achieved in the present work by exploiting the marching cubes algorithm (section \ref{sec:Marching cubes algorithm}). The result of the reconstructed topology is shown in Figure \ref{fig:reconstruction_overview}c, where a small ``bridge'' is created between the particles around the merging point.

\subsection{Marching cubes algorithm}
\label{sec:Marching cubes algorithm}
\begin{figure}[ht]
    \centering
    \includegraphics[]{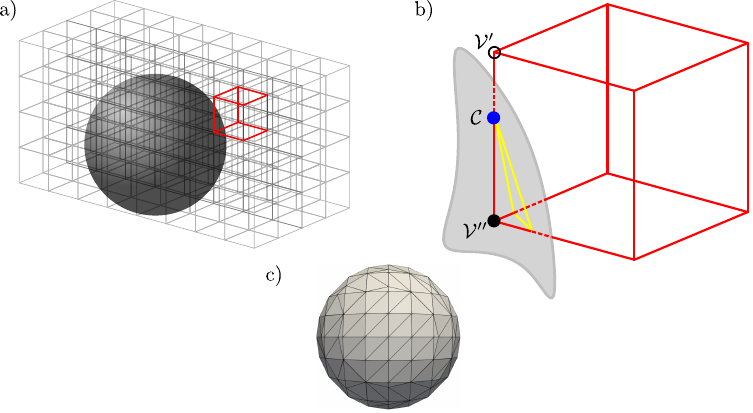}
    \caption{Main steps of marching cubes triangulation. Auxiliary Cartesian mesh (a). Compute intersection points $\mathcal{C}$ from linear interpolation between values at vertices $\mathcal{V}'$ (outside the interface), $\mathcal{V}''$ (inside the interface) and assemble triangular elements (b). Final triangulated body (c).}
    \label{fig:marchingcubes_review}
\end{figure}

Marching cubes is a robust and fast algorithm for surface reconstruction from an indicator function field and was originally proposed by \citet{Lorensen1987} for medical applications (e.g., computed tomography, magnetic resonance). Let $f(x,y,z) = 0$ be a (known) continuous surface (a sphere in the example of Figure \ref{fig:marchingcubes_review}) that we wish to approximate with a discrete triangular mesh. 
The first step consists of creating an equidistant Cartesian mesh (independent of the Eulerian grid used for the governing equations) that fully overlaps the region occupied by the surface $f(x,y,z) = 0$ (Figure \ref{fig:marchingcubes_review}a). For each vertex $\mathcal{V}$ of the grid with coordinates $(x_\mathcal{V}, y_\mathcal{V}, z_\mathcal{V})$, it is possible to compute the value of $f$ at the corresponding location, i.e., $f_\mathcal{V} = f(x_\mathcal{V}, y_\mathcal{V}, z_\mathcal{V})$. The intersections between the surface and the edges of the Cartesian grid can be simply obtained by looping over each edge of the mesh (delimited by the generic vertices $\mathcal{V}'$ and $\mathcal{V}''$) and check for the product $f_{\mathcal{V}'} f_{\mathcal{V}''}$. When such a quantity is negative, it means that there is a change in the sign of $f$ between $\mathcal{V}'$ and $\mathcal{V}''$ and an intersection (i.e., where $f=0$) occurs accordingly in between (Figure \ref{fig:marchingcubes_review}b). The coordinates of the intersection point $(\mathcal{C})$ are readily obtained with a linear interpolation scheme (high-order approaches are obviously possible \citep{Evrard2019}):
\begin{equation}
\label{eq:linear_interpolation}
    \overline{\mathcal{V}' \mathcal{C}} = \left( \frac{f_{\mathcal{V}'}}{f_{\mathcal{V}'} - f_{\mathcal{V}''}} \right) \overline{\mathcal{V}' \mathcal{V}''}
\end{equation}
where $\overline{\mathcal{V}' \mathcal{C}}$ is the distance between the first edge vertex and the intersection point. This procedure is repeated in a loop over the whole mesh and, once all the edges have been traversed, the intersection points are connected to form a triangular mesh (Figure \ref{fig:marchingcubes_review}c). This procedure ensures that each vertex $\mathcal{C}$ is always the same (i.e., same coordinates) for the four adjacent cubes that share edge $\mathcal{V}' \mathcal{V}''$, since the values $f_{\mathcal{V}'}, f_{\mathcal{V}''}$ are the same. However, this is not sufficient to ensure a full continuity of the triangular front, as there could be configurations where a triangular edge is not shared by two adjacent triangles. This scenario leads to the formation of holes in the reconstructed front and will be discussed in section \ref{sec:Marching cubes ambiguities}. {In the case of a free-surface flow, all the peripheral edges are connected to one triangle only and, therefore, such elements must be excluded when performing checks for holes in the reconstructed interface.}  {In the limit case of $f = 0$ at one cube vertex, a triangle with collapsed vertices and null area would be reconstructed. To avoid such a degenerate configuration, we check for the value of $f$ at each cube vertex and, if $|f_\mathcal{V}| < 10^{-6}$, we increase $f_\mathcal{V}$ by $\pm 10^{-3}$, where the sign is the same as the sign of $f_\mathcal{V}$. Such an increase in the value of $f$ guarantees that the methodology is robust and capable of handling complex interface reconstructions.} An advantage of the marching cubes algorithm consists of the ease of parallelisation because only local information is needed. Indeed, only the values of $f$ at the vertices are needed and the reconstruction of the triangles is performed locally within each cell without any communication required from the neighbouring elements. This is true as long as the intersection points are computed with a first-order scheme (e.g., linear interpolation of equation \eqref{eq:linear_interpolation}), whilst high-order approaches would require larger stencil and read access to data from the neighbouring cells.  

\begin{figure}[ht]
    \centering
    \includegraphics[]{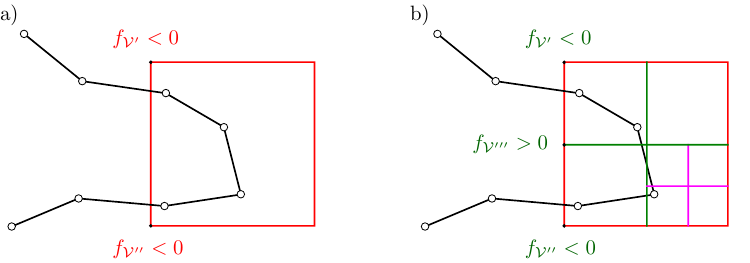}
    \caption{Reconstruction of a highly-curved interface. Under-resolved (a) and refined (b) grids.}
    \label{fig:under_resolution}
\end{figure}

The procedure discussed above illustrates the discretization of a surface for which the equation is known. In the case of topology changes in two-phase flows, the resulting shape of the interface is not known a-priori and the analytical formula that describes the shape cannot be obtained. {However, in order to reconstruct the interface after topology changes, the analytical representation of its shape is not necessary.} As is illustrated in Figure \ref{fig:reconstruction_overview}, it is possible to derive the shape of the interface by computing the iso-surface $S_{\mathcal{I} = 0.5}$ of the indicator function. The Poisson equation, equation \eqref{eq:poisson}, is solved on the Cartesian grid used for front reconstruction with marching cubes and the values of $\mathcal{I}$ are stored at the cell centres. Such values are then interpolated at the vertices $\mathcal{V}$ by means of a simple arithmetic average, which leads to the definition of the vertex scalar field $\mathcal{I}_\mathcal{V}$. At this point, we can apply the procedure illustrated above for the reconstruction of the analytical function $f(x, y, z)$, with the only difference that the intersection points belong to the set of points $\mathcal{C} \colon \mathcal{I}_\mathcal{C} = 0.5$, i.e., $f = \mathcal{I} - 0.5$. {Reconstructing the iso-surface for $\mathcal{I} = 0.5$ is therefore equivalent to searching for the points belonging to $f = 0$.} {In the following we assume a point to be inside (outside) the interface if the corresponding indicator value is $\mathcal{I} > 0.5$ ($\mathcal{I} < 0.5$).} {The indicator function is not the only possible choice to reconstruct the interface. The same marching cubes procedure can be used to find the iso-surface $\phi = 0$, where $\phi$ is a signed distance function that can be derived from the Lagrangian front \citep{Shin2009}.} One distinguished feature of the marching cubes reconstruction method is that it is possible to reach a high level of accuracy in the discretization of the iso-surface. The Cartesian mesh used for the reconstruction is not necessarily the same as the one used for the solution of the Navier-Stokes equations. This allows us to decouple the reconstruction procedure from the solution of the governing equations, which can then be solved on any kind of mesh, e.g., unstructured grids. It is therefore possible to use a very fine grid for marching cubes without affecting the computational cost of the governing equations. This feature is particularly relevant when the interface exhibits a strong curvature. The under-resolution of such regions would lead to cases where multiple intersections occur for one or more edges of the marching cubes grid, leading to loss of information in this area, since only one intersection per edge can be captured by the algorithm. An example (2D) of an under-resolved reconstruction is shown in Figure \ref{fig:under_resolution}a, where the front intersects twice with the $\mathcal{V}' \mathcal{V}''$ edge.  
Since $f_{\mathcal{V}'} f_{\mathcal{V}''} > 0$, the algorithm returns no intersection. This issue is readily fixed by refining the marching cubes grid (Figure \ref{fig:under_resolution}b). The edge between vertices $\mathcal{V}'$ and $\mathcal{V}''$ is now split in two smaller edges with opposite boundary values (i.e., $f_{\mathcal{V}'} f_{\mathcal{V}'''} < 0$ and $f_{\mathcal{V}'''} f_{\mathcal{V}''} < 0$) with one intersection each. Due to the very strong curvature of the proposed example, the bottom right cell of the refined mesh is under-resolved too and would require further refinement (e.g., the purple grid in Figure \ref{fig:under_resolution}b). {The choice of the marching cubes grid resolution is therefore a compromise between accuracy and computational cost. Interfaces with highly-curved regions are naturally more demanding in terms of resolution. Once the appropriate size of the marching cubes grid is chosen, the same spacing is applied everywhere, i.e., the reconstruction grid is a uniform Cartesian mesh. It is possible to combine adaptive mesh refinement techniques with marching cubes (see, for example, the work of \citet{Shu1995}), but this approach is not implemented in the present work.}
The use of a different, highly-resolved grid clearly comes with the cost of solving twice equation \eqref{eq:poisson}, since the indicator function is still needed on the Eulerian grid for the one-fluid formulation of the Navier-Stokes equations. It is finally reminded here that Peskin spreading for the indicator gradients $\nabla \mathcal{I}$ is based (in 3D) on a $5 \times 5 \times 5$ stencil. When refining the mesh for marching cubes, it could happen that the front triangles before topology changes are too large compared to the cell size of the auxiliary mesh and their contributions (in terms of $\nabla \mathcal{I}$) reach cells not included in the spreading stencil. To avoid this, the Lagrangian front is first refined to be consistent with Peskin spreading.

\begin{figure}[ht]
    \centering
    \includegraphics[]{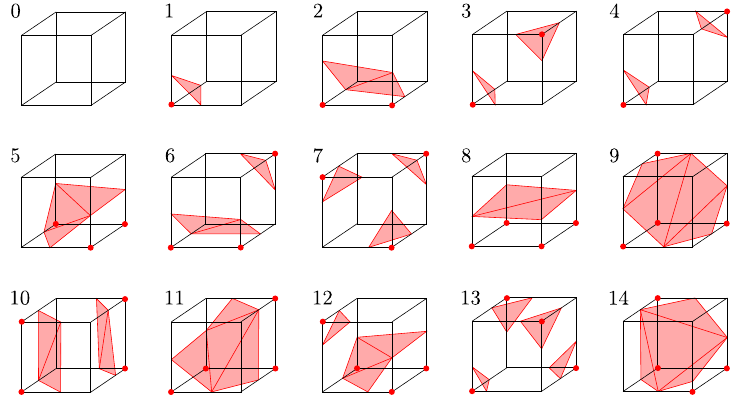}
    \caption{Lookup table of the unique 15 cases as identified in \cite{Lorensen1987}. The vertices where $\mathcal{I} > 0.5$ are identified with a red circle.}
    \label{fig:lookup_table}
\end{figure}

\begin{figure}[ht]
    \centering
    \includegraphics[]{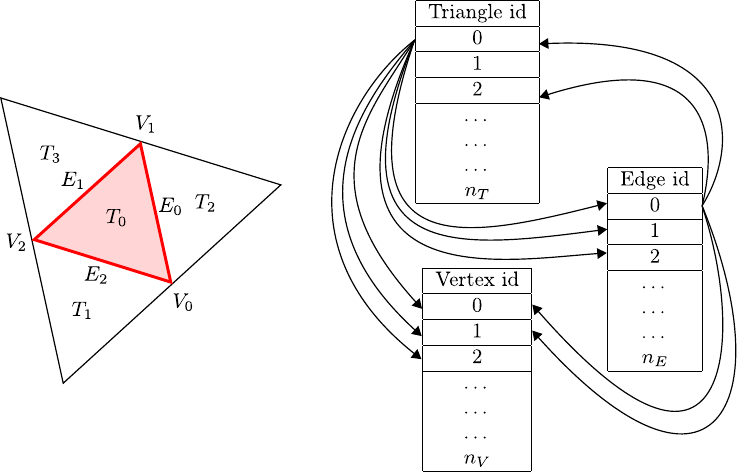}
    \caption{Connectivity of the triangular front. Each triangle has links to its vertices and edges. Each edge is connected to the corresponding two vertices and triangles that share the edge itself.}
    \label{fig:mftl}
\end{figure}

The marching cubes algorithm is implemented by listing all the possible cubes configurations in a lookup table. Since each vertex can be either inside $(\mathcal{I}_\mathcal{V} > 0.5)$ or outside $(\mathcal{I}_\mathcal{V} < 0.5)$ the iso-surface, the total amount of possible configurations is $2^8 = 256$. However, the authors in \cite{Lorensen1987} make use of symmetries to show that the number of entries in the lookup table can be reduced to only 15 unique cases (Figure \ref{fig:lookup_table}). Since all the possible topological configurations are pre-computed, the marching cubes algorithm is extremely efficient. 
Once the specific cube configuration is identified, the number of triangles (up to four) and their arrangement within the cell is known a-priori from the lookup table. For each edge that owns a triangle vertex, the coordinates of the vertex are computed with equation \eqref{eq:linear_interpolation} and stored in a hash list. In a similar way, lists are also created for edges and triangles. The connectivity among these elements is then ensured by linking each edge to the corresponding two vertices and two triangles that share it. Finally, for each element in the list of triangles, the links to the three vertices and edges owned by the triangle are populated. The connectivity among the front elements is sketched in Figure \ref{fig:mftl}.
{At the beginning of a simulation, each single body (i.e., a connected Lagrangian front representing, for example, a bubble or droplet) is initialised with the data structure shown in Figure \ref{fig:mftl}. To distinguish among multiple bodies, a unique id is assigned to each of them as soon as they are initialised.}

\subsubsection{Marching cubes ambiguities}
\label{sec:Marching cubes ambiguities}
\begin{figure}[ht]
    \centering
    \includegraphics[]{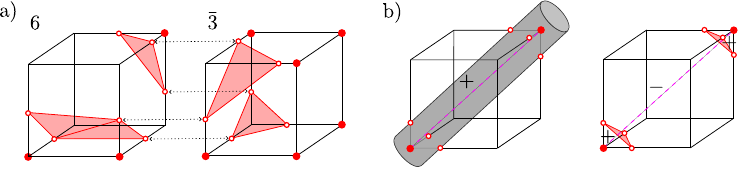}
    \caption{(a) Face ambiguity that leads to the formation of a hole on the face shared between cube 6 (left) and the complementary configuration of case 3 (right). (b) Internal ambiguity for the reconstruction of a cylinder (left). Cube configuration 4 (right) only takes into account one topological reconstruction, which is not consistent with the original surface.}
    \label{fig:ambiguities_review}
\end{figure}

A known issue of the marching cubes algorithm is the existence of ambiguities in the arrangement of triangles for some of the 15 cases \cite{Nielson1991, Nagae1993, Montani1994, Natarajan1994, Chernyaev1995, Newman2006}. Such ambiguities are generally distinguished based on whether they occur on the face (face ambiguities) or inside of the cubes (internal ambiguities). Face ambiguities affect each case where all the four edges of a face are cut by the iso-surface $S_{\mathcal{I} = 0.5}$. This happens when, for each set of two adjacent vertices $\left( \mathcal{V}', \mathcal{V}'' \right)$, one is inside and the other one outside the interface, i.e., $\left( \mathcal{I}_{\mathcal{V}'} - 0.5 \right) \left( \mathcal{I}_{\mathcal{V}''} - 0.5 \right) < 0$. Such ambiguity can lead to a hole in the reconstructed front on the ambiguous face and affects cases 3, 6, 7, 10, 12 and 13 of the lookup table. An example is shown in Figure \ref{fig:ambiguities_review}a, where the inconsistency occurs on the face shared by cube 6 and the complementary configuration of case 3 (referred to as $\bar{3}$). 

The other type of ambiguities arises inside the cube whenever the reconstructed front is topologically inconsistent. An example of such inconsistency is shown in Figure \ref{fig:ambiguities_review}b for the reconstruction of a circular cylinder. The configuration under investigation is case 4 from the original lookup table, which leads to two disconnected triangles. Marching cubes, without knowledge of the ambiguity, clearly produces a wrong result as it can be easily seen by looking at the dotted line that connects the lower-left-front vertex with the upper-right-rear one, where any point that belongs to this line is inside the original surface (i.e., $f>0$, indicated with a `$+$' symbol). However, in the reconstructed configuration, the same line intersects the triangular front twice, leading to a set of points (between the two triangles) that lies outside the iso-surface (`$-$' symbol). Since marching cubes processes each cube independently, a natural approximation of the indicator function $\left( \bar{\mathcal{I}} \right)$ follows from the trilinear interpolation:
\begin{align}
\label{eq:trilinear}
    \begin{split}
        \bar{\mathcal{I}} &= \mathcal{I}_{\mathcal{V}^{000}} (1 - q) (1 - r) (1 - s)\\
        &+ \mathcal{I}_{\mathcal{V}^{100}} q (1 - r) (1 - s)\\
        &+ \mathcal{I}_{\mathcal{V}^{010}} (1 - q) r (1 - s)\\
        &+ \mathcal{I}_{\mathcal{V}^{110}} q r (1 - s)\\
        &+ \mathcal{I}_{\mathcal{V}^{001}} (1 - q) (1 - r) s\\
        &+ \mathcal{I}_{\mathcal{V}^{101}} q (1 - r) s\\
        &+ \mathcal{I}_{\mathcal{V}^{011}} (1 - q) r s\\
        &+ \mathcal{I}_{\mathcal{V}^{111}} q r s
    \end{split}
\end{align}
where $(q,r,s)$ is the cube coordinate system and $\mathcal{I}_{\mathcal{V}^{000, 100, \ldots 111}}$ are the corresponding vertex values. {Internal ambiguities may arise in marching cubes because the cases listed in the original lookup table only consider vertex values and, for the same set of indicator values at the cube vertices, multiple topologies are possible, i.e., the shape inside the cube is not uniquely defined. However, the topological correct front is the only configuration that has the same topology of the one resulting from the trilinear interpolation of the indicator function \citep{Chernyaev1995}. In the ambiguous cases, the lookup table only provides one possible topology, which does not necessarily match the one from the trilinear interpolation, leading to the presence of ambiguities.} Internal ambiguities can affect cases 3, 4, 6, 7, 10, 12 and 13, i.e., the same inconsistent configurations for face ambiguities (plus cube 4). 

Several works have addressed the issue of ambiguities in marching cubes. Face inconsistencies are treated in \citet{Nielson1991} and \citet{Montani1994}. The former introduces an asymptotic decider algorithm to chose the correct edge pattern on ambiguous faces, whilst the latter proposes new complementary configurations for the ambiguous cases in order to avoid the formation of holes. The asymptotic decider algorithm introduced by \citet{Nielson1991} is based on the assumption of bilinear variation of $\mathcal{I}$ over a plane (the two-dimensional equivalent of equation \eqref{eq:trilinear}) and ensures that the reconstructed front is continuous $\left( C^0 \right)$. This work is extended in \citet{Natarajan1994} and \citet{Chernyaev1995} to include internal ambiguities as well. \citet{Natarajan1994} checks for the value of the interpolated indicator function at the saddle point of the trilinear surface to decide whether two regions with the same sign of $\mathcal{I}$ are connected or disconnected. \citet{Chernyaev1995} uses bilinear variation of $\mathcal{I}$ on internal planes, parallel to one of the cube faces, to identify the correct topological configuration and proposes a new version of marching cubes with a lookup table of 33 cases with no ambiguities. This work is further optimised and extended in \cite{Lewiner2003, Custodio2013, Vega2019}. For a review on the marching cubes algorithm and the several techniques that can be adopted to avoid inconsistencies, the reader is referred to the work of \citet{Newman2006}.

Face and internal ambiguities lead to non-continuous and topologically inconsistent triangular fronts that do not have a physical meaning. It is important to note that the ambiguous cases are the configurations that contain multiple disconnected triangles (i.e., cases 3, 4, 6, 7, 10, 12 and 13 of the lookup table). Face ambiguities that lead to holes in the interface are never encountered in any of our simulations. {In the following we assume to deal with closed interfaces and, according to the front topology shown in Figure \ref{fig:mftl}, each edge is linked to the two adjacent triangles that share the edge itself.} In the case of a hole, there would be some edges linked to one triangle only (Figure \ref{fig:ambiguities_review}a). {After each reconstruction, we check for edges connectivity and, if a triangle is missing, the simulation is interrupted. It is noted here that the marching cubes algorithm does not distinguish between closed or free surfaces and can be applied to both cases. In the case of free surfaces, the peripheral triangles must be excluded from this check.} Internal ambiguities do not lead to crashes and can actually occur in our methodology. However, the occurrence of ambiguous configurations is generally very limited, owing to the way the indicator function is computed on the Eulerian grid. The spreading of the indicator gradient $\nabla \mathcal{I}$ (equation \eqref{eq:poisson}) produces a scalar field $\mathcal{I}$ that varies smoothly between the limit values $(0 < \mathcal{I} < 1)$ across approximately three layers of cells (Figure \ref{fig:reconstruction_overview}c). The result resembles the indicator function field obtained with an algebraic volume of fluid method and the corresponding interface (represented by the iso-surface $S_{\mathcal{I} = 0.5}$) generally does not have two separate parts in the same Eulerian cell. {This happens since the indicator function only provides the relative amount of the primary phase volume, but not the number of interfaces that are present inside the cell. Therefore, even for cases that before front reconstruction have multiple interfaces inside the same marching cubes cell, after solving equation \eqref{eq:poisson} for $\mathcal{I}$ and interpolating the values at the cube vertices, the corresponding configurations from the lookup table are generally non-ambiguous cases.} Table \ref{table:frequency_ambiguities} anticipates the occurrence of the ambiguous configurations for the validation cases presented in section \ref{sec:validation}. 
\begin{table}[ht]
\caption{Frequencies (\%) of marching cubes cases. Configuration $C0$, where the cube is empty, is not relevant and has been excluded from the statistics. The ambiguous cases are highlighted in red.}
\centering
 \begin{tabular}{l >{\raggedright\arraybackslash}p{2.5cm} >{\raggedright\arraybackslash}p{2.5cm} >{\raggedright\arraybackslash}p{2.5cm} >{\raggedright\arraybackslash}p{2.5cm} >{\raggedright\arraybackslash}p{2.5cm}} 
 \hline
 Config. 
 &Single sphere (\ref{sec:Volume error due to front reconstruction}) 
 &Droplet breakup in shear flow(\ref{sec:Droplet breakup in simple shear flow}) &Droplet-droplet collision (\ref{sec:Droplet-droplet collision}) 
 &Rising bubble (peripheral breakup - \ref{sec:Breakup of a rising bubble})
 &Rising bubble (central breakup - \ref{sec:Breakup of a rising bubble})  \\ [0.5ex] 
 \hline
 $C1$ &27.2 &19.5 &26.5 &29.05 &28.6\\
 $C2$ &34.0 &38.4 &33.5 &29.89 &31.8\\
 {$C3$} &{0} &{0.0005} &{0} &{0.53} &{0}\\
 {$C4$} &{0} &{0} &{0} &{0.04} &{0}\\
 $C5$ &16.7 &15.8 &18.5 &18.63 &18.7\\
 {$C6$} &{0} &{0} &{0} &{0.33} &{0}\\
 {$C7$} &{0} &{0} &{0} &{0} &{0}\\
 $C8$ &16.9 &24.4 &17.5 &16.49 &15.9\\
 $C9$ &5.2 &1.9 &4.0 &4.95 &5.0\\
 {$C10$} &{0} &{0} &{0} &{0.02} &{0}\\
 $C11$ &0 &0 &0 &0.03 &0\\
 {$C12$} &{0} &{0} &{0} &{0.01} &{0}\\
 {$C13$} &{0} &{0} &{0} &{0} &{0}\\
 $C14$ &0 &0 &0 &0.03 &0\\
 \hline
 \end{tabular}
 \label{table:frequency_ambiguities}
\end{table}
For each of the 256 cases that can occur during the reconstruction procedure in each cube, the corresponding configuration from the lookup table is found and its frequency monitored during the simulation (the cases that are either completely inside or outside the interface are excluded). The results show that the ambiguous configurations (highlighted with a red font colour) represent {a very limited number of cases} among all the possible cases (only the rising bubble with peripheral breakup exhibits some occurrences, but less than 1\% globally). It is finally reminded that an ambiguous case does not necessarily leads to a topologically inconsistent reconstruction, i.e., not all the ambiguous cubes in Table \ref{table:frequency_ambiguities} necessarily represent a wrong interface. From the above considerations, it follows that the error induced by topologically inconsistent reconstruction has a marginal effect on the evolution of the entire interface. {Despite the different approach, a similar conclusion applies to the LCRM/LFRM methods. The issue of ambiguities is treated by splitting the reconstruction cubes into tetrahedral cells, which return only one possible topology. However, an ambiguity still persists as the direction of the splitting is arbitrary and different choices lead to different topologies (e.g., it can prevent or induce topology changes). To mitigate this effect, the orientation of the tetrahedral cells is periodically switched. Regardless of the specific orientation choice, these methodologies have proven to be effective in handling topology changes, proving that ambiguities do not have a central role in this type of simulations.}

\subsection{Detection of topology changes}
\label{sec:Detection of topology changes}
{The marching cubes based reconstruction algorithm is only needed for topology changes. It is therefore important to guarantee that only the bodies undergoing breakup/coalescence are actually reconstructed. From an algorithmic point of view, before generating the marching cubes grid, it is necessary to detect and flag all the front objects that experience breakup and/or coalescence.  This represents one of the main differences of the proposed methodology compared to other works that exploit grid-based reconstruction approaches, such as LCRM \citep{Shin2002, Shin2005, Shin2007a, Yoon2010} and LFRM \citep{Shin2011}. This aspect has two advantages:} not all the tracked interfaces need to be reconstructed, thus saving accuracy by avoiding unnecessary reconstruction errors, and marching cubes is executed only at the time steps where breakup and/or coalescence are detected. The first point brings a significant reduction in the computational cost of the reconstruction procedure since the grid used for marching cubes can potentially be much smaller than the computational domain occupied by the entire set of fronts. The second advantage has a strong impact on the cost as well, since the occurrences of breakup/coalescence events are generally smaller than the total number of time steps of the simulation. Therefore, marching cubes is executed only a few times during the simulation, as opposed to front reconstruction operations in LCRM/LFRM, which occur periodically with a rough estimate of one reconstruction every 100 time steps \citep{Shin2011}.

\begin{figure}[ht]
    \centering
    \includegraphics[]{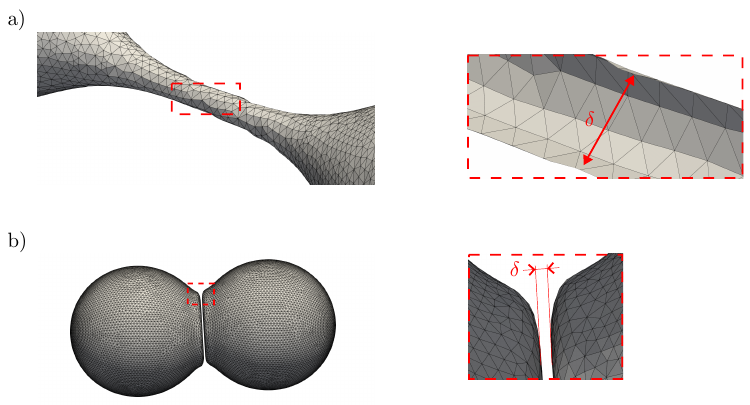}
    \caption{Detection of breakup of a fluid ligament (a) and coalescence of two droplets (b). The identification of topology-changing bodies is based on the minimum distance $\delta$ between two triangular elements with opposite normal vectors.}
    \label{fig:BreakupAndCoalescence}
\end{figure}

{Detection of topology changes essentially includes two types of interfacial configurations:} breakup (Figure \ref{fig:BreakupAndCoalescence}a), where a single body develops a thin filament that can lead to the formation of two separated interfaces, and coalescence (Figure \ref{fig:BreakupAndCoalescence}b), where multiple interfaces close to each other can be merged into a single object. 
In both cases, detection is based on a threshold distance ($\delta$ in Figure \ref{fig:BreakupAndCoalescence}). In principle, for each triangle of a single front, it is possible to loop over the other triangles of the same body (breakup) or of another object (coalescence) and check the distance between each pair of triangle centres. If such a distance is below the threshold value $\delta$, we check the orientation of the two triangles and, if the normal vectors have opposite direction (i.e., $\mathbf{n}_{T_1} \cdot \mathbf{n}_{T_2} < 0$, where $T_1$ and $T_2$ refer to two generic triangles), the object (breakup) or pair of objects (coalescence) are flagged as topology-changing bodies. It can be immediately seen that the computational cost of such an algorithm is $\mathcal{O}(n_T^2)$ for breakup, where $n_T$ is the number of triangles of the single object, and $\mathcal{O}(n_{1,T} n_{2,T})$ for coalescence, where the subscripts 1 and 2 refer to two different bodies with ids 1 and 2, respectively. For interfaces with a large number of triangles, the execution of the detection algorithm can be expensive. To reduce the cost of such operation, we use a hierarchical structure of bounding boxes, which allows to perform fast rejection tests for a large group of elements without looping over all the triangles of a front. 

\subsubsection{Bounding boxes}
\begin{figure}[ht]
    \centering
    \includegraphics[]{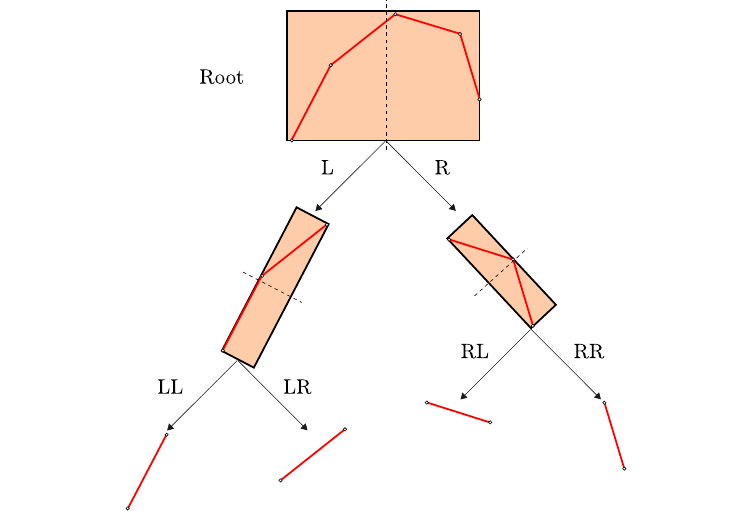}
    \caption{Bounding boxes tree structure for a 2D front. Each box is split along the largest principal axis and contains the front elements from the parent node whose centre lies on the same side of the cut (e.g., box L owns the two elements of the root node with centres on the left side of the cut).}
    \label{fig:BboxSketch}
\end{figure}

{We adopt the tree structure of oriented bounding boxes presented by \citet{Gottschalk1996}}. An example of bounding boxes for a 2D front is shown in Figure \ref{fig:BboxSketch}. 
The first box (root) encompasses the entire set of triangles (segment lines in 2D) and is the largest box in the tree. {The root box must contain the object entirely and, at the same time, fit the triangulated surface as tightly as possible. This is achieved by computing the covariance matrix $\mathbf{C}$ defined as:}
\begin{equation}
        C_{i,j} = \frac{1}{3n_T} \sum_{k = 0}^{n_T} (\bar{x}_{0,i}^k \bar{x}_{0,j}^k + \bar{x}_{1,i}^k \bar{x}_{1,j}^k + \bar{x}_{2,i}^k \bar{x}_{2,j}^k)
\end{equation}
{where $n_T$ is the number of triangles. The vector $\bar{\mathbf{x}}_i$ is computed as $\bar{\mathbf{x}}_i = \mathbf{x}_i - \xi$, with the coordinates of the $i-\text{th}$ vertex ($i = 0,1,2$) of the generic triangle represented by $\mathbf{x}_i$. The mean value $\xi$ follows from:}
\begin{equation}
        \xi = \frac{1}{3n_T} \sum_{k = 0}^{n_T} (\mathbf{x}_0 + \mathbf{x}_1 + \mathbf{x}_2)
\end{equation}
{Two out of the three eigenvectors of the covariance matrix $\mathbf{C}$ represent the maximum and minimum variance and tend to align the bounding box with the triangulated object. The box is therefore oriented with the orthogonal basis given by the eigenvectors of $\mathbf{C}$, and the size is determined by the intersections between the vectors and the body, ensuring that the object is entirely contained within the bounding box.} The next tree level is obtained by splitting the root box along the largest principal axis and consists of two children boxes with two elements each for the example shown in Figure \ref{fig:BboxSketch}. A generic box contains all the front elements owned by the parent node whose centre lies on the same side of the cut. The procedure is repeated until only one element per box is left. The last boxes of each path in the tree are also referred to as leaf nodes. Each box can be identified by its path from the root level, which consists of a sequence of L (left) and R (right) characters, according to the direction of each split. When the bounding box tree is created, the coordinates of the centre and dimensions of each box as well as the list of triangular elements contained within are computed and stored in the hierarchical structure.  

\begin{figure}[ht]
    \centering
    \includegraphics[]{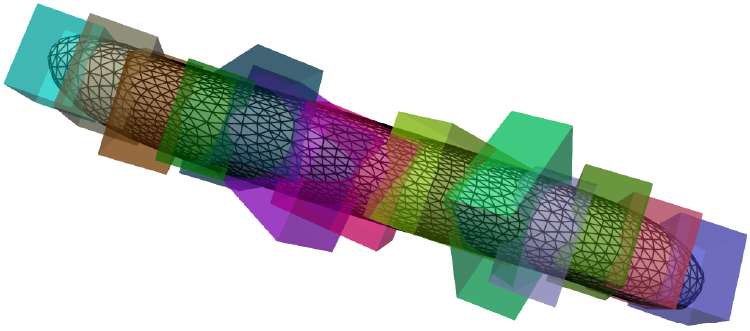}
    \caption{Bounding boxes used for detection of breakup of a droplet in a shear flow. The distance between pairs of triangle centres is tested only within the corresponding box. The threshold value $\delta$ has been increased (compared to realistic values) and some of the bounding boxes removed to improve the readability of the figure.}
    \label{fig:BboxShearDroplet}
\end{figure}

When looking for breakup, once we reach a bounding box with one of the principal dimension smaller than $2 \delta$, the distance between each pair of triangles contained within the same box is computed and, if it is smaller than $\delta$ and the scalar product between the normal vectors is negative, the body is flagged and we interrupt the loop. If the smallest distance is still larger than $\delta$, we keep iterating over the remaining bounding boxes of similar size. This approach reduces the computational cost of the algorithm, since the list of triangles to loop over is significantly smaller compared to the entire front. A 3D representation of the bounding boxes with one of the dimensions smaller than $2 \delta$ is shown in Figure \ref{fig:BboxShearDroplet} for a deforming droplet in a planar shear flow.
{It is finally noted that there might be a configuration in which a pair of triangles with centre to centre distance smaller than $\delta$ is not detected by the breakup algorithm if the two elements belong to two adjacent bounding boxes. However, the breakup event is generally anticipated by the formation of a thin filament with several triangles close to each other, and it is necessary to identify only one pair of these elements to trigger the breakup of the filament. In front tracking simulations of realistic two-phase systems, the filaments that undergo breakup must be resolved with enough triangular elements. Using bounding boxes with size $2 \delta$ is expected to be accurate enough to guarantee that topology changes are executed consistently with the chosen breakup criterion.}

\begin{figure}[ht]
    \centering
    \includegraphics[]{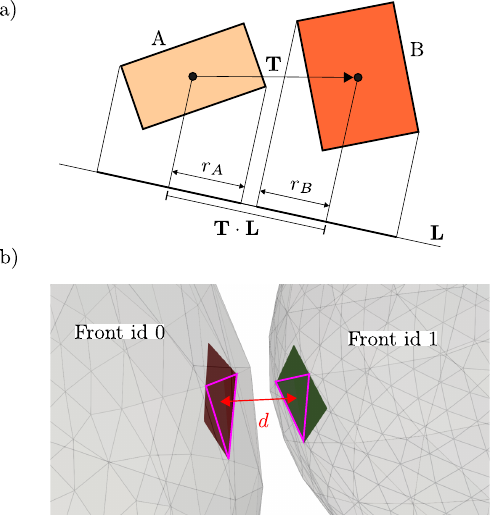}
    \caption{(a) Overlapping test for two bounding boxes (A, B) based on inequality \eqref{eq:overlap_inequality}. (b) The overlapping test returns two bounding boxes leaves (red and green for fronts 0 and 1 respectively). The distance between the triangles contained inside the boxes is checked against the threshold value. If $d \leq \delta$, the two bodies are flagged as merging.}
    \label{fig:BboxOverlap}
\end{figure}

A similar approach is used for the detection of merging interfaces, where two bodies, which we identify with ids 0 and 1, approach each other and merge into a single continuous front (coalescence). The algorithm is based on the work of \citet{Gottschalk1996}, who developed a methodology to compute the intersections between two bodies. We first introduce the original algorithm (for intersections) and then discuss the necessary modifications for coalescence. The first step consists of creating the bounding boxes trees for both bodies and check if the two root boxes overlap.
Two boxes (A and B in Figure \ref{fig:BboxOverlap}a) do not overlap if the following condition is true:
\begin{equation}
\label{eq:overlap_inequality}
    | \mathbf{T} \cdot \mathbf{L} | > |r_A| + |r_B|
\end{equation}
The above inequality is tested by performing 15 tests with different directions $\mathbf{L}$ and, if in all cases the inequality is not satisfied, the boxes overlap each other. {The choice of the 15 tests is based on the separating axis theorem for a three-dimensional space, which follows as a corollary of the separating plane theorem (the reader is referred to \citet{Gottschalk2000} for more details).} If there is an overlap, some of the triangles contained inside the root box of body 0 {may intersect some} of the elements owned by body 1; otherwise the two boxes are completely disjointed with no intersections. In the event of an overlap, the test is repeated by navigating through the bounding box tree of body 0 until we reach one of the leaf nodes that still overlaps the root box of the other body (the boxes that do not overlap are discarded). At this point, we loop over the tree structure of body 1 discarding all the boxes that do not overlap the leaf node. At the end of this step, we are left with one leaf node of body 0 and one leaf (or more)  of body 1. The intersection between these triangles can then be computed. The whole algorithm is repeated until all the possible configurations of overlapping leaf nodes from both bodies are found. In our case, instead of intersections, we look for pairs of triangles with a centre to centre distance below the threshold value $\delta$. The same algorithm illustrated above can be used by relaxing the overlap condition:
\begin{equation}
\label{eq:merging_inequality}
    | \mathbf{T} \cdot \mathbf{L} | > |r_A| + |r_B| + 5\delta
\end{equation}
In this way, we do not restrict the search to overlapping boxes but we keep into consideration the boxes with a distance between elements compatible with the length $\delta$ (the value $5 \delta$ is used as a safety factor). The output of the algorithm consists of a set of triangle pairs with a distance between their centres potentially smaller than $\delta$, as illustrated in Figure \ref{fig:BboxOverlap}b where also the corresponding leaf boxes are shown. We then check each triangle pair and, if $d < \delta$, both bodies are flagged for topology changes. {In the case where the two bodies are far apart from each others, inequality \ref{eq:merging_inequality} is satisfied (for at least one of the 15 tests) and the searching algorithm is immediately aborted, since no further traversal of the children bounding boxes is required.}

Both algorithms for detection of breakup and coalescence considerably decrease the size of triangle sets that need to be tested, thus reducing significantly the overall computational cost. {A front can be flagged for breakup, coalescence (with, at least, another object) or both. The specific criterion that flags the body is not relevant, since the marching cubes algorithm relies on the indicator function only and the object is entirely reconstructed, regardless of the specific topology change event.} {It is finally noted that combining classical remeshing with the detection algorithm allows us to avoid undesired topology changes. This is especially relevant for breakup, where two triangular elements (of the same front) with opposite normals are allowed to coexist inside the same cell, provided the threshold $\delta$ is small enough. The triangular elements can still be regularised via remeshing without enforcing topology changes. On the other hand, if the front regularisation is based on the reconstruction of the interface, a breakup would occur as soon as the distance between the elements is smaller than the cell size.}

\subsubsection{Assignment of unique ids}
\label{sec:Assignment of unique ids}
\begin{figure}[ht]
    \centering
    \includegraphics[]{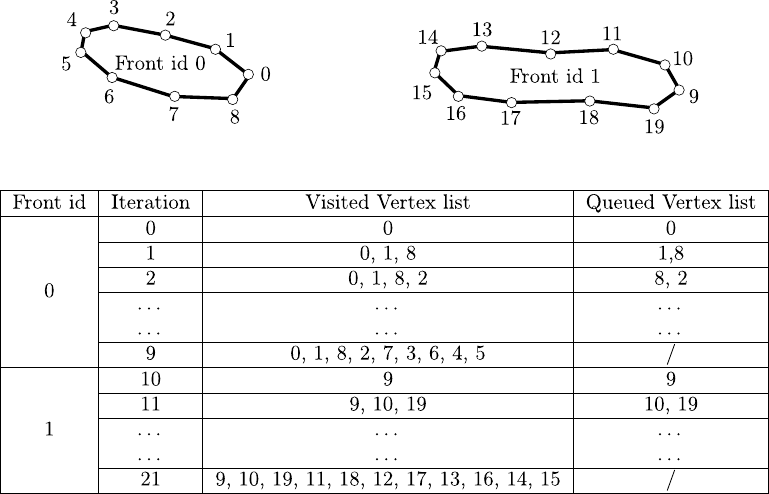}
    \caption{Breadth First Search algorithm for the assignment of unique ids to bodies after reconstruction.}
    \label{fig:BreadthFirstSearch}
\end{figure}

After all the bodies are tested for topology changes, we temporarily remove those that are not flagged and we proceed with front reconstruction with marching cubes. The main steps are:
\begin{enumerate}
    \item Compute the minimum and maximum $x, y, z$ coordinates from all the flagged bodies. Build a Cartesian mesh that spans from $(x_\text{min}, y_\text{min}, z_\text{min})$ to $(x_\text{max}, y_\text{max}, z_\text{max})$. 
    \item Compute the indicator function on the newly created Cartesian mesh by solving equation \eqref{eq:poisson}.
    \item Interpolate the indicator function at the Cartesian vertices $\mathcal{V}$.
    \item Execute marching cubes.
\end{enumerate}
As explained in section \ref{sec:Marching cubes algorithm}, the output of marching cubes consists of three lists of connected elements, i.e., vertices, edges and triangles. During the reconstruction step, it is not possible to identify the elements that form each single body. For example, if we reconstruct a droplet that breaks into two children droplets, the front that we obtain after marching cubes has a unique id, although two different ids should be used (one per droplet after breakup). The assignment of unique ids to each individual object is performed after the reconstruction step, using the breadth first search algorithm, as illustrated in Figure \ref{fig:BreadthFirstSearch}.
In this example we obtain two bodies after reconstruction that we want to distinguish with front ids 0 and 1. The first step of the algorithm consists of creating two lists (visited vertex and queued vertex lists) and populate both with the first vertex in the list obtained from marching cubes (i.e., vertex 0). Exploiting the connectivity between vertices, edges and triangles, we look for the vertices directly connected to vertex 0 (i.e., those that share one edge - in this example these are vertex 1 and 8) and add them to both lists. When no more connected vertices are found, vertex 0 is deleted from the queued list. The algorithm is repeated for each element in the queued list until this list is left empty. All the traversed vertices belong to the same front and are assigned to the same id. If there are still vertices that have not been visited (e.g., vertex 9), the procedure is repeated until all the vertices are traversed. After the last iteration of the algorithm is executed, we obtain unique ids for each reconstructed body. Since the bodies that were not flagged for topology changes already had their id, we ensure that the ids assigned by the breadth first search algorithm are not duplicates. The computational cost of the breadth search algorithm is $\mathcal{O}(n_V + n_E)$, where $n_V, n_E$ are the number of vertices and edges, respectively.    

When the assignment of unique ids is concluded, the non-flagged bodies (i.e., the non topology-changing bodies) are inserted back in the list of tracked interfaces. The indicator function is then solved on the Eulerian mesh and used for the solution of the Navier-Stokes equations.

\subsection{Mesh quality of reconstructed fronts}
\label{sec:Mesh quality of reconstructed fronts}
\begin{figure}[ht]
    \centering
    \includegraphics[]{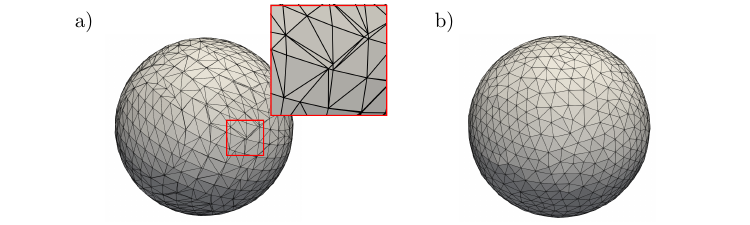}
    \caption{Reconstructed sphere with marching cubes before (a) and after (b) remeshing.}
    \label{fig:remeshing}
\end{figure}

A good mesh quality of the Lagrangian front is necessary to ensure accurate results in terms of curvature and normal vector computation and, therefore, surface tension effects. However, the mesh obtained from the marching cubes reconstruction step is usually characterised by a non-even distribution of the vertices, which leads to the formation of highly-skewed triangles and non-uniform area distribution (i.e., large area ratio between two adjacent triangles). An example of reconstructed mesh for a spherical interface is shown in Figure \ref{fig:remeshing}a.
In order to ensure a good quality of the front mesh, each reconstructed body is remeshed after the marching cubes step. Such remeshing consists of smoothing the distribution of vertices (through a relaxation scheme applied along the tangential direction \citep{Botsch2011}) and performing operations on triangle edges (e.g., split, merge and swap) as usually done in front tracking \citep{Gorges2022}. The result after remeshing is shown in Figure \ref{fig:remeshing}b, where a good quality front with a uniform element distribution is obtained.

\subsection{Parallel implementation}
\begin{figure}[ht]
        \centering
        \includegraphics[]{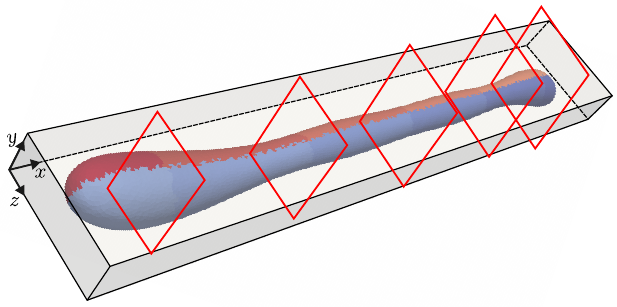}
        \caption{Domain decomposition of a Lagrangian data structure. Each colour represents the elements owned by a different processor. The red planes show the cuts along the $x-$direction for the decomposition of the orthogonal box around the object.}
        \label{fig:DropletParallelisation}
    \end{figure}

{The parallelisation of front tracking algorithms is notoriously difficult, but extremely important when large-scale systems are simulated. The proposed framework uses domain decomposition and message-passing to parallelize the operations on the Lagrangian front, including the marching cubes based reconstruction. Since a detailed discussion on parallelisation is beyond the scope of this work, only the main steps are discussed below. The generic Lagrangian structure shown in Figure \ref{fig:mftl} is distributed across the processors. Each processor stores the lists of elements (i.e., vertices, edges and triangles) that it owns. A vertex is owned by a processor if its coordinates are inside the corresponding sub-domain. Since edges and triangles may cross the domain decomposition boundaries, it is important to ensure that the adjacent processors have access to these elements, by either owning them or as ghost elements. A processor stores all the edges and triangles for which it owns at least one vertex. Each edge and triangle are linked to an ordered list of two and three vertices, respectively, and they are owned by a processor if the same processor owns the first vertex of the list. Otherwise, they are stored as ghost elements. Domain decomposition is performed in a Cartesian way (i.e., with cuts along the $x-$, $y-$ and $z-$directions) of an orthogonal box that fits the Lagrangian front. Such decomposition is decoupled from the parallel management of the Navier-Stokes grid, allowing for flexibility in the choice of the computational mesh (e.g., structured or unstructured). An example of domain decomposition is shown in Figure \ref{fig:DropletParallelisation}.} 

\begin{figure}[ht]
        \centering
        \includegraphics[]{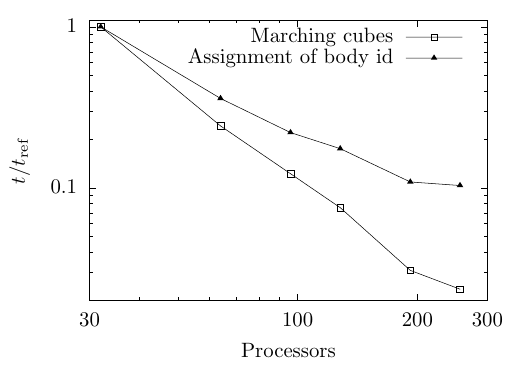}
        \caption{Parallel scalability benchmark for the marching cubes and assignment of unique ids routines. Time is made non-dimensional with the corresponding execution time ($t_\mathrm{ref}$) from the case with the lowest number of processors.}
        \label{fig:plot_scalability}
    \end{figure}

{The main steps of the reconstruction procedure are also parallel. The Poisson equation \eqref{eq:poisson} is solved with a parallel biconjugate gradient stabilized method (BiCGSTAB) incorporated in the open source PETSc library \citep{Balay1997, petsc-user-ref}. The absolute tolerance is set to $10^{-8}$ and the solver generally converges in less than 200 iterations. Once the indicator function is interpolated at the cubes vertices, marching cubes can be easily parallelised, since the reconstruction is performed within each cube and no additional information from the neighbours is needed. Detection of topology changes, as well as assignment of unique ids, are performed in parallel, following the decomposition of the Lagrangian front. A scalability benchmark is reported in Figure \ref{fig:plot_scalability} for the marching cubes and id assignment routines, where multiple spherical droplets are initialised in a cubic domain and the total number of triangular elements is approximately $3 \times 10^5$.} 
{Both algorithms scale up to almost 300 processors. It is finally noted that the Cartesian domain decomposition is not load balanced and the overall efficiency depends on the specific configuration.}

\subsection{Overview of the algorithm for topology changes}
\begin{algorithm}[H]
\label{alg:marching_cubes}
\caption{Front tracking algorithm with topology changes}
\SetAlgoLined
\KwIn{Initialisation of front tracking structures: list of bodies and front connectivity (vertices, edges and triangles)}
\vspace{0.5em}  

\tcc{Start time loop}
\While{$t < t_{\text{end}}$}{
    Velocity interpolation from the Eulerian mesh to the Lagrangian vertices\;
    Front advection (equation \eqref{eq:front_advection})\;
    Front remeshing\;
    \tcc{Topology changes}
    Breakup/Coalescence detection (section \ref{sec:Detection of topology changes})\;
    \If{topology changes are detected}{
    Remove the non topology-changing bodies from the fronts list\;
    Create a Cartesian mesh for marching cubes that overlaps the bodies\;
    Solve the indicator function on the new mesh (equation \eqref{eq:poisson})\;
    Interpolate the indicator function from the cell centres to the vertices\;
    Execute marching cubes\;
    Front remeshing\;
    Assign a unique id to the reconstructed bodies (section \ref{sec:Assignment of unique ids})\;
    Move back the non topology-changing bodies into the fronts list\;
    }
    \tcc{One-fluid formulation of the governing equations}
    Solve the indicator function (equation \eqref{eq:poisson}) on the Eulerian mesh used for the discretization of the Navier-Stokes equations\;
    Compute surface tension term (equation \eqref{eq:surface_tension})\;
    Update fluid properties $\rho, \mu$ (equation \eqref{eq:density})\;
    Solve Navier-Stokes equations\;
}
\end{algorithm}

\section{Validation}
\label{sec:validation}
In this section, the presented methodology is tested against several benchmarks that include breakup and coalescence of fluid particles (e.g., droplet, bubbles). The two-phase system under consideration is a disperse flow and we use the subscript $d, c$ for the disperse and continuous phases, respectively. For example, in the case of a bubble (gas) rising in a large tank filled with liquid, $\rho_\mathrm{d}$ is the density of the gaseous phase, whilst the liquid has density $\rho_\mathrm{c}$.  

\subsection{Volume error due to front reconstruction}
\label{sec:Volume error due to front reconstruction}
\begin{figure}[ht]
    \centering
    \includegraphics[]{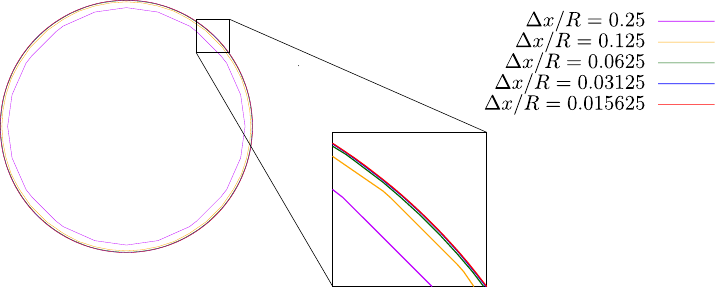}
    \caption{Slices of reconstructed spheres at different resolutions. $\Delta x$ is the grid size used for marching cubes, whilst the Eulerian mesh remains unchanged.}
    \label{fig:sphere_reconstruction}
\end{figure}

\begin{figure}[ht]
    \centering
    \includegraphics[]{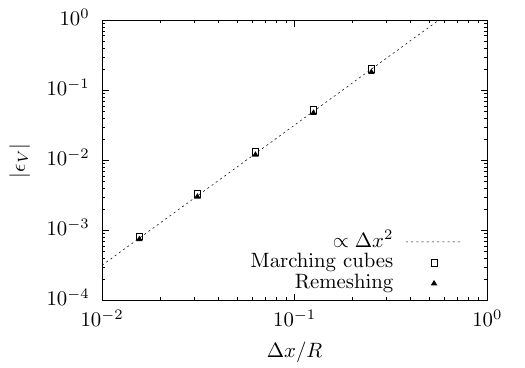}
    \caption{Relative volume error for the reconstruction of a single sphere. The volume is computed with equation \eqref{eq:volume} and evaluated right after the marching cubes step and after the following remeshing operation.}
    \label{fig:plot_volume_error_R1}
\end{figure}

In this benchmark, we quantify the error and its order of convergence when the front is reconstructed with marching cubes. We initialise the Lagrangian front with a unit size sphere (i.e., the diameter is $D = 1$) and reconstruct the interface with the procedure discussed above. We perform five test cases where we progressively refine the Cartesian mesh (with size $\Delta x$) used for reconstruction, whilst the Eulerian grid for the solution of the governing equations stays always the same. Since the focus is on the reconstruction error, we perform one iteration only. The qualitative results are presented in Figure \ref{fig:sphere_reconstruction}, where slices of the reconstructed sphere are compared for resolutions varying between $0.015625 R < \Delta x < 0.25 R$ ($R$ is the radius of the sphere). 
The reconstructed shape converges with the refinement of the Cartesian grid and the effect of decreasing $\Delta x$ becomes almost negligible at large resolutions, e.g., $\Delta x < 0.0625 R$. The reconstruction error is based on the volume of the sphere and is computed as:
\begin{equation}
    \epsilon_V = \frac{V_\text{reconstructed} - V_\text{ref}}{V_\text{ref}}
\end{equation}
where $V_\text{ref}$ is the volume of the Lagrangian front as is initialised, whilst $V_\text{reconstructed}$ is the volume of the object returned by marching cubes. {The volume of the triangulated front is computed as:}
\begin{equation}
\label{eq:volume}
        V = \sum_{k = 0}^{n_T} \left[ \frac{1}{6} N_x^k (x_{0,x}^k + x_{1,x}^k + x_{2,x}^k) \right]
\end{equation}
{where $n_T$ is the number of triangles. For the generic $k-\text{th}$ triangle, $N_x^k$ is the x-component of the cross-product of the edge vectors $(\mathbf{x}_1 - \mathbf{x}_0)$, $(\mathbf{x}_2 - \mathbf{x}_0)$, whilst $(\mathbf{x}_0, \mathbf{x}_1, \mathbf{x}_2)$ are the coordinates of the vertices. The formula inside the summation operator returns the volume between the $k-\text{th}$ Lagrangian triangle and its projection onto the $x = 0$ plane.} The corresponding order of convergence is shown in Figure \ref{fig:plot_volume_error_R1}, {where the volume is monitored right after the marching cubes step as well as after the remeshing step that follows (see section \ref{sec:Mesh quality of reconstructed fronts}).}
The plot shows second-order convergence for both quantities, consistently with the numerical schemes involved in the marching cubes reconstruction procedure. The solution depends on the accuracy with which the indicator function is obtained on the Cartesian mesh used for reconstruction. This involves two steps (see algorithm \ref{alg:marching_cubes}):
\begin{enumerate}
    \item Lagrangian to Eulerian spreading of the indicator gradient via Peskin weighting approach.
    \item Solution of the Poisson equation.
\end{enumerate}
The first step involves the weighting defined in equation \eqref{eq:PeskinEquation}, which is second-order accurate when interpolating smooth functions \citep{Peskin2003}. The second step consists of solving an elliptic equation with a standard second-order centred finite difference scheme. The marching cubes error is therefore consistent with the errors introduced in the above steps.

\subsection{Droplet breakup in simple shear flow}
\label{sec:Droplet breakup in simple shear flow}
\begin{figure}[ht]
    \centering
    \includegraphics[]{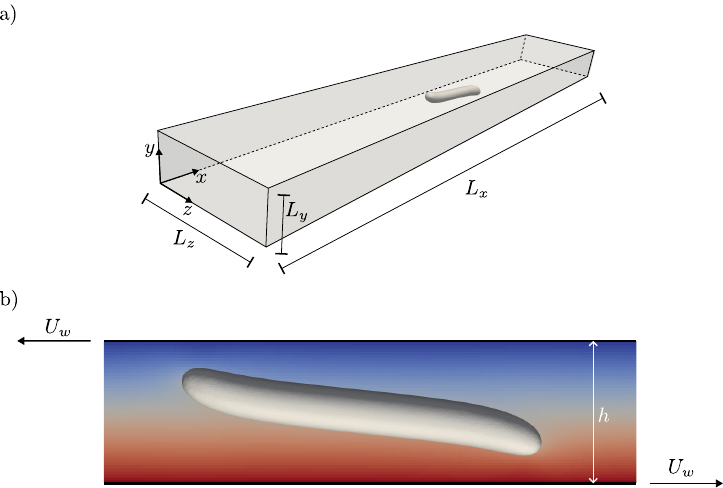}
    \caption{Setup of a droplet in a shear flow.}
    \label{fig:sketchDropletShear}
\end{figure}

This test case models the breakup of a droplet in a planar shear flow. The computational domain consists of a channel of length $L_x$ and cross section with dimensions $L_y \times L_z$ (Figure \ref{fig:sketchDropletShear})a. 
The extension along the $y$-direction is the height of the channel and is referred to with the symbol $h$. The upper $(y = h)$ and bottom ($y = 0$) walls move with opposite velocity but same magnitude $U_w$, i.e., $-U_w \mathbf{\hat{i}}$ and $U_w \mathbf{\hat{i}}$ respectively. Symmetry boundary conditions are applied to the side walls ($z = 0$, $z = L_z$), whilst the faces at $x = 0$, $x = L_x$ are periodic. For laminar flows, the velocity field within the channel has a steady-state solution, with analytical form:
\begin{equation}
    \mathbf{u} = U_\mathrm{w} \left( 1 - \frac{2}{h}y \right) \mathbf{\hat{i}}
\end{equation}
The contour of the $x$-component of $\mathbf{u}$ on a plane is shown in Figure \ref{fig:sketchDropletShear}b. The droplet is initialised as a sphere with diameter $D$ and placed at the centre of the channel, which has dimensions $L_x \times L_y \times L_z = 24D \times 10D/7 \times 4D$ and a diameter to height ratio of $D/h = 0.7$. 

The problem is governed by four non-dimensional numbers. The properties of the two-phase system are compared by the density and viscosity ratios:
\begin{equation}
\label{eq:density_visc_ratio}
    \rho_\mathrm{r} = \frac{\rho_\mathrm{d}}{\rho_\mathrm{c}}, \quad \mu_\mathrm{r} = \frac{\mu_\mathrm{d}}{\mu_\mathrm{c}}
\end{equation}
The Reynolds number is computed as:
\begin{equation}
    \mathrm{Re} = \frac{\rho_\mathrm{c} \dot{\gamma} R^2}{\mu_\mathrm{c}}
\end{equation}
where $R$ is the (initial) droplet radius and $\dot{\gamma}$ is the shear rate inside the channel, i.e., $\dot{\gamma} = 2 U_w / h$. The ratio between viscous and surface tension forces is given by the {capillary} number:
\begin{equation}
    \mathrm{Ca} = \frac{\mu_\mathrm{c} \dot{\gamma} R}{\sigma}
\end{equation}

We reproduce one of the examples introduced in \citet{Sibillo2006}, where the authors perform experiments at low $\mathrm{Re}$, in a two-phase system with $\rho_\mathrm{r} = \mu_\mathrm{r} = 1$ and {capillary} number equal to $\mathrm{Ca} = 0.46$. The exact Reynolds number is not reported, but we found the value $\mathrm{Re} = 0.5$ a good compromise between the numerical cost of the simulation (the lower the $\mathrm{Re}$, the longer it takes to reach the breakup point) and the modelling of a flow regime dominated by viscous forces. The problem of a droplet deforming in a simple shear flow has been investigated in the literature for a long time \citep{Hinch1980, Grace1982, Bruijn1989, Guido2001} and the {capillary} number has been used to identify the critical state where breakup occurs, i.e., for $\mathrm{Ca} > \mathrm{Ca}_\text{cr}$. For the two-phase system under consideration, the critical {capillary} number in an unbounded flow (i.e., $D/h << 1$) is $\mathrm{Ca}_\text{cr} = 0.43$. In the present case, there is a confinement effect (due to the large diameter to height ratio, $D/h = 0.7$) that leads to a slight increase in the critical value. Nonetheless, we are still in a super-critical regime $(\mathrm{Ca} > \mathrm{Ca}_\text{cr})$ where breakup is observed \citep{Sibillo2006}. The Eulerian mesh has a uniform resolution $(\Delta x)$ with approximately 33 cells per droplet diameter, whilst $\Delta t$ is limited by the capillary time step constraint \citep{Denner2015}. {A mesh sensitivity analysis is performed with resolutions up to 45 cells per diameter and shows that no significant improvements are achieved for grid sizes $\Delta x < D/33$.} The Cartesian mesh for marching cubes has a grid spacing of $\Delta x/2$ {and the threshold value for breakup is set to the same resolution, i.e., $\delta = \Delta x / 2$.} The characteristic parameters that define the problem are reported in Table \ref{table:DropletShear}.
\begin{table}[ht]
\caption{Non-dimensional governing parameters for a droplet in simple shear flow.}
\centering
 \begin{tabular}{c c c c c c c >{\raggedright\arraybackslash}p{2.5cm} >{\raggedright\arraybackslash}p{2.5cm} c} 
 \hline
 $\rho_\mathrm{r}$ &$\mu_\mathrm{r}$ &$\mathrm{Re}$ &$\mathrm{Ca}$ &$L_x$ &$L_y$ &$L_z$ &Eulerian grid $(\Delta x)$ &Marching cubes grid &$\delta$\\ [0.5ex] 
 \hline
 1 &1 &0.5 &0.46 &$24D$ &$10D/7$ &$4D$ &$D/33$ &$\Delta x /2$ &{$\Delta x / 2$} \\
 \hline
 \end{tabular}
 \label{table:DropletShear}
\end{table}

\begin{figure}[ht]
    \centering
    \includegraphics[]{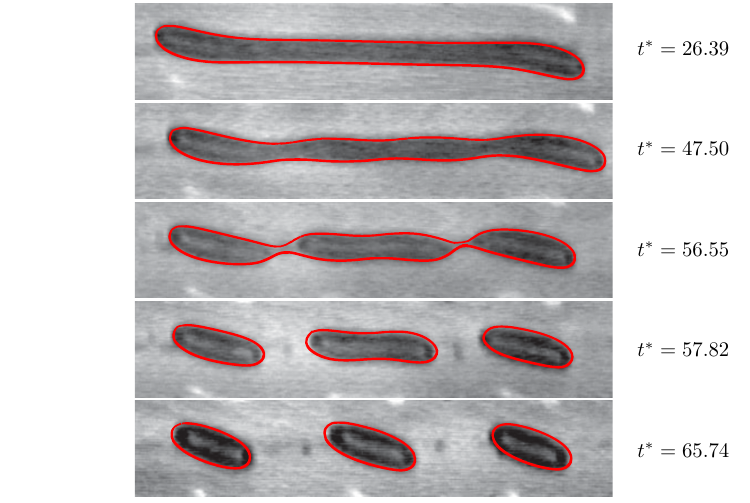}
    \caption{Droplet breakup in a simple shear flow with $\mathrm{Ca} = 0.46$, $\mathrm{Re} = 0.5$, $\rho_\mathrm{r} = \mu_\mathrm{r} = 1$ and $D/h = 0.7$. The background is from the experiments of \citet{Sibillo2006}, the red profiles represent the simulation. Time is made non dimensional with $t^* = t \Dot{\gamma}$. Figure adapted with permission from \citet{Sibillo2006}. Copyrighted by the American Physical Society.}
    \label{fig:Sibillo_fig3_top}
\end{figure}

\begin{figure}
    \centering
    \includegraphics[]{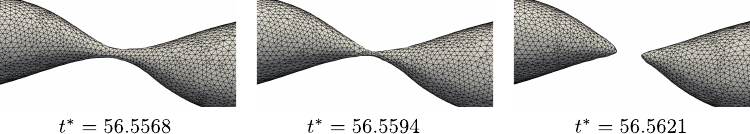}
    \caption{Details of the Lagrangian mesh near one of the breakup points.}
    \label{fig:mesh_detail}
\end{figure}

The time evolution of the deforming droplet is shown in Figure \ref{fig:Sibillo_fig3_top}, where time is made non-dimensional with the reference time $1/\dot{\gamma}$, i.e., $t^* = t \Dot{\gamma}$. 
The snapshots in the background are taken from the experiments of \citet{Sibillo2006}, whilst the red profiles represent the numerical results obtained with the proposed methodology. The droplet develops into an elongated body that resembles an ellipsoid with the two ends that point towards the closest moving wall. The orientation angle between the flow direction and the main dimension of the ellipsoid decreases over time and becomes almost null (i.e., the droplet is aligned horizontally) for $t^* \approx 26$. The maximum elongation is reached at $t^* \approx 47$, where interfacial instabilities, in the form of waves, appear in the central part of the droplet. The competition between surface tension and shear forces lead to the formation of three main parts connected by two thin liquid filaments ($t^* = 56.55$) that then break up. The algorithm for topology changes (section \ref{sec:Detection of topology changes}) detects breakup between $56.55 < t^* < 57.82$. After breakup, we observe three children droplets with the lateral ones oriented along the direction of the shear flow and the central one aligned horizontally. At a later time, the central droplet aligns to the flow direction as well. As observed in \citet{Sibillo2006}, the central droplet is larger than the lateral ones, which is a consequence of the confinement effect. Overall, a good comparison is observed between experiments and numerical simulations. The front tracking methodology is able to accurately reproduce both elongation and orientation of the droplet, whilst the topology changes algorithm effectively captures the breakup point. Details of the Lagrangian mesh before and after breakup are shown in Figure \ref{fig:mesh_detail}, where a good quality of the mesh is maintained during the breakup process, owing to the remeshing of the front after the execution of the marching cubes algorithm. 

\subsubsection{Effect of varying the marching cubes grid resolution and breakup threshold}
\label{sec:Effect of varying the marching cubes grid resolution and breakup threshold}
\begin{figure}[ht]
        \centering
        \includegraphics[]{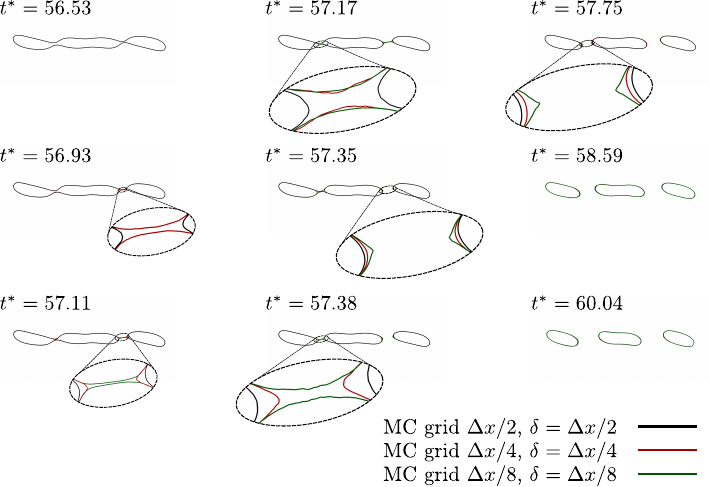}
        \caption{Sensitivity analysis to the marching cubes grid resolution and breakup threshold for a droplet in simple shear flow. The profiles correspond to slices of the Lagrangian front on the $x-y$ plane.}
        \label{fig:DropletShearFlow_mesh_sensitivity}
\end{figure}

{The choice of the resolution of the reconstruction grid, as well as the breakup and coalescence thresholds, are critical aspects of the proposed methodology and are expected to influence the results. In this section we investigate the effect of varying these parameters for the breakup of a droplet in simple shear flow. The cases for the sensitivity analysis are reported in Table \ref{table:DropletShearMeshSensitivity}, where the first case corresponds to the one already discussed in section \ref{sec:Droplet breakup in simple shear flow}.}
\begin{table}[ht]
    \caption{Setup for the sensitivity analysis to the marching cubes grid resolution and breakup threshold for a droplet in simple shear flow.}
    \centering
        \begin{tabular}{c c c c c c c >{\raggedright\arraybackslash}p{2.5cm} >{\raggedright\arraybackslash}p{2.5cm} c} 
        \hline
        $\rho_\mathrm{r}$ &$\mu_\mathrm{r}$ &$\mathrm{Re}$ &$\mathrm{Ca}$ &$L_x$ &$L_y$ &$L_z$ &Eulerian grid $(\Delta x)$ &Marching cubes (MC) grid &$\delta$\\ [0.5ex] 
        \hline
        1 &1 &0.5 &0.46 &$24D$ &$10D/7$ &$4D$ &$D/33$ &$\Delta x /2$ &$\Delta x / 2$ \\
        1 &1 &0.5 &0.46 &$24D$ &$10D/7$ &$4D$ &$D/33$ &$\Delta x /4$ &$\Delta x / 4$ \\
        1 &1 &0.5 &0.46 &$24D$ &$10D/7$ &$4D$ &$D/33$ &$\Delta x /8$ &$\Delta x / 8$ \\
        \hline
        \end{tabular}
        \label{table:DropletShearMeshSensitivity}
\end{table}
{Compared to this simulation, the second and third cases employ a factor of 2 and 4, respectively, to refine the marching cubes grid and breakup threshold $\delta$. The results are shown in Figure \ref{fig:DropletShearFlow_mesh_sensitivity}, where several slices on the $x-y$ plane are compared.}
{As expected, decreasing the threshold $\delta$ leads to a delay in the breakup event. This can be easily observed by monitoring the detachment of the right droplet satellite, which occurs at $t^* \approx 56.93$, $t^* \approx 57.11$ and $t^* \approx 57.35$ for $\delta = \Delta x /2$, $\delta = \Delta x /4$ and $\delta = \Delta x /8$, respectively. A similar mechanism is observed for the left satellite. The main consequence consists of a longer elongation of the three detached droplets as $\delta$ decreases. However, such differences are relatively minor and the global dynamics of the droplet, in terms of number of satellites, their orientation and shape, is not significantly affected by the value of $\delta$. By decreasing the size of the marching cubes grid, the volume error introduced by the reconstruction procedure decreases as well (with a second-order convergence rate, as shown in Figure \ref{fig:plot_volume_error_R1}). However, the larger error that we introduce in the case where the resolution of the MC grid is the lowest (i.e., $\Delta x /2$), does not affect significantly the accuracy of the solution, since the reconstruction step is executed only when it is strictly necessary. For many disperse flow simulations, such events occur with a low frequency and accurate results can be obtained without refining the MC grid excessively compared to the Eulerian grid.}

\subsection{Droplet-droplet collision}
\label{sec:Droplet-droplet collision}
\begin{figure}[ht]
    \centering
    \includegraphics[]{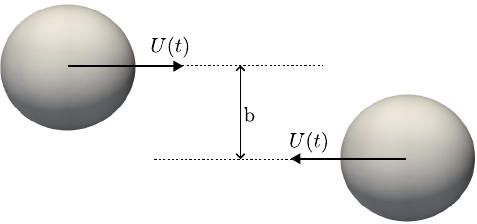}
    \caption{Setup of two colliding droplets.}
    \label{fig:sketchDropletCollision}
\end{figure}

This validation case models the collision of two liquid droplets, which results in the merging of the interfaces \citep{Qian1997, Finotello2018}. Two equal spherical droplets (with diameter $D$) are accelerated towards each other until the interfaces collide. A sketch of the problem setup is shown in Figure \ref{fig:sketchDropletCollision}.
The droplets travel along the horizontal direction and the vertical distance between their centre is $b$. The velocity $U(t)$ is time dependent, since both particles are accelerated from rest. In the following, we will omit the time dependence to refer to the velocity at the collision point, i.e., $U(\Tilde{t}) = U$, where $\Tilde{t}$ is the time instant where collision occurs.

The two-phase system under consideration is defined by the density and viscosity ratios (equation \eqref{eq:density_visc_ratio}). The problem is further described by the following non-dimensional numbers. The impact parameter is used to describe off-centre collisions and is defined as:
\begin{equation}
    \mathrm{I} = \frac{b}{D}
\end{equation}
For $I = 0$, the droplet centres are aligned to the direction of relative motion and a head-on collision takes place. The ratio between inertial and surface tension forces is quantified by the Weber number:
\begin{equation}
    \mathrm{We} = \frac{\rho_\mathrm{d} D \left( U_\text{rel} \right)^2}{\sigma}
\end{equation}
where $U_\text{rel} = 2 U$ is the relative velocity between the droplets. Viscous and surface tension effects are compared in the Ohnesorge number:
\begin{equation}
    \mathrm{Oh} = \frac{\mu_\mathrm{d}}{\sqrt{\rho_\mathrm{d} D \sigma}}
\end{equation}
Here we replicate one of the experiments performed in the work of \citet{Finotello2018}, where a milk solution with total solid (TS) content of 20\% is used for the droplets. For solutions below 30\% TS, the milk phase can be treated as a Newtonian fluid. The computational domain is a box with dimensions $15D \times 10D \times 10D$ (the droplets are accelerated along the $x$-direction) and the Eulerian mesh has a resolution ($\Delta x$) of 32 cells per droplet diameter. Such a large domain is chosen to avoid confinement effects due to the presence of walls, which are treated with symmetric boundary conditions. {A mesh sensitivity analysis is performed with resolutions up to 42 cells per droplets diameter and shows that no significant improvements are achieved for grid sizes $\Delta x < D/33$.} The Cartesian mesh for marching cubes has the same resolution as the Eulerian one, whilst the threshold distance for topology changes detection is $\delta = \Delta x / 10$. A summary of the characteristic parameters are reported in Table \ref{table:DropletCollision}.
\begin{table}[ht]
\caption{Non-dimensional governing parameters for droplet-droplet collision.}
\centering
 \begin{tabular}{c c c c c c c c >{\raggedright\arraybackslash}p{2.5cm} >{\raggedright\arraybackslash}p{2.5cm} c} 
 \hline
 $\rho_\mathrm{r}$ &$\mu_\mathrm{r}$ &$\mathrm{I}$ &$\mathrm{We}$ &$\mathrm{Oh}$ &$L_x$ &$L_y$ &$L_z$ &Eulerian grid $(\Delta x)$ &Marching cubes grid &$\delta$\\ [0.5ex] 
 \hline
 864.6 &237.2 &0.05 &59.2 &0.022 &$15D$ &$10D$ &$10D$ &$\Delta x = D/33$ &$\Delta x$ &$\Delta x / 10$\\
 \hline
 \end{tabular}
 \label{table:DropletCollision}
\end{table}
A small value of the impact parameter $I$ is used here, which corresponds to an (almost) head-on collision.

\begin{figure}[ht]
    \centering
    \includegraphics[]{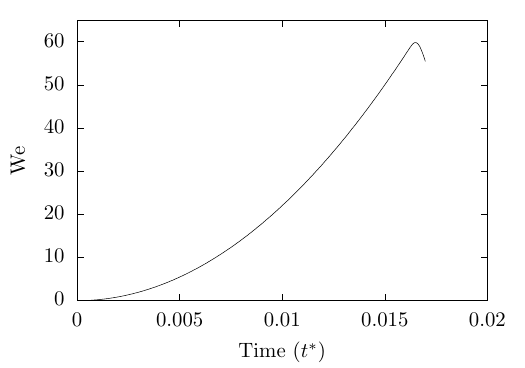}
    \caption{Plot of Weber number vs time.}
    \label{fig:Weber}
\end{figure}

In our numerical model, we use a body force along the $x$-component to accelerate the droplets and initialise their motion \citep{Nobari1996, Razizadeh2018}:
\begin{equation}
    f_x = C (\rho - \rho_\mathrm{c})
\end{equation}
The value of $f_x$ is non null only inside the disperse phase (i.e., the droplets), whilst it takes a null value in the rest of the domain. The constant $C$ is adjusted here to set the acceleration of the droplets so that the relative velocity ($2U$) at the collision point gives the desired Weber number. The time evolution of $\mathrm{We}$ is shown in Figure \ref{fig:Weber}, which shows a quadratic profile (being $U \propto t$ for a constant acceleration) until the maximum value of $\mathrm{We} \approx 60$ is reached. 
The deceleration that follows is due to the proximity effect between the droplets when they are very close to each other. In this work we consider as the reference Weber number for collision the maximum value reached by $\mathrm{We}$ before deceleration. The body force $f_x$ is switched off as soon as coalescence takes place. The initial distance {(measured along the horizontal direction)} between the droplet centres is $4D$.

\begin{figure}[ht]
    \centering
    \includegraphics[]{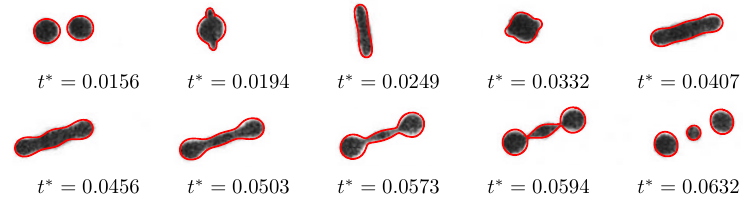}
    \caption{Droplet-droplet collision with $\mathrm{We} = 60$, $\mathrm{Oh} = 0.022$ and $b/D = 0.05$. The background is from the experiments of \citet{Finotello2018}, the red profiles represent the simulation. Time is made non-dimensional with $t^* = t \mu_\mathrm{d} / \left( \rho_\mathrm{d} D \right)$. Figure adapted from \citet{Finotello2018} (CC BY 4.0).}
    \label{fig:Finotello2018_fig8d}
\end{figure}

The comparison between the work of \citet{Finotello2018} and our simulations is shown in Figure \ref{fig:Finotello2018_fig8d}, where the background pictures come from the experimental investigation and the red profiles are the numerical results.
After the droplets merge, a transient dynamics with continuous changes in the shape characterises the motion. The effect of the offset $(b)$ between the two original centres is visible after the merging at $t^* = 0.0194$, with the right side slightly shifted upwards with respect to the left one. The droplet then deforms into a shape that resembles a disk $(t^* = 0.0249)$, before changing its orientation and relaxing into an elongated ellipsoidal body. Interfacial undulations appear on the droplet at $t^* = 0.0456$, which lead to the formation of a liquid bridge between the two spherical caps. A contraction mechanism can be observed for $t^* > 0.0503$, where the spherical ends begin to move towards each other. The amount of kinetic energy put into the system before collision is enough to break the liquid bridge in two points near the end caps, creating a small satellite droplet in between. Such a collision regime is called reflexive separation (due to the breakup after collision) and the good agreement between experiments and simulations shows that the proposed methodology is able to resolve the physics of the problem.

\subsubsection{Effect of varying the breakup and coalescence threshold}
\begin{figure}[ht]
        \centering
        \includegraphics[]{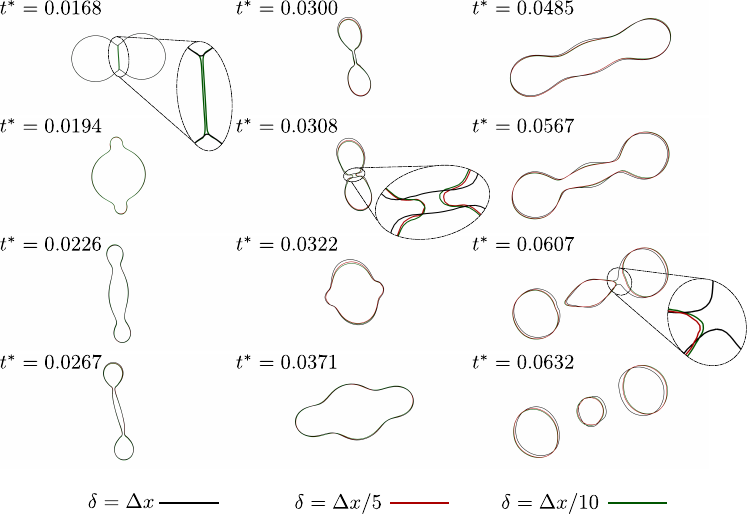}
        \caption{Sensitivity analysis to the coalescence/breakup threshold for a droplet-droplet collision. The profiles correspond to slices of the Lagrangian front on plane of the relative motion.}
        \label{fig:DropletCollisionFlow_mesh_sensitivity}
\end{figure}

{We investigate here the effect of varying the threshold parameter $\delta$ for topology changes in the complex scenario of colliding droplets, where both coalescence and breakup take place. The marching cubes grid is kept constant, with a resolution equal to $\Delta x$. The cases for the sensitivity analysis are reported in Table \ref{table:DropletCollisionMeshSensitivity}, where the last case corresponds to the one already discussed in section \ref{sec:Droplet-droplet collision}.} 
\begin{table}[ht]
    \caption{Setup for the sensitivity analysis to the coalescence/breakup threshold for a droplet-droplet collision.}
    \centering
        \begin{tabular}{c c c c c c c c >{\raggedright\arraybackslash}p{2.5cm} >{\raggedright\arraybackslash}p{2.5cm} c} 
        \hline
        $\rho_\mathrm{r}$ &$\mu_\mathrm{r}$ &$\mathrm{I}$ &$\mathrm{We}$ &$\mathrm{Oh}$ &$L_x$ &$L_y$ &$L_z$ &Eulerian grid $(\Delta x)$ &Marching cubes (MC) grid &$\delta$\\ [0.5ex] 
        \hline
        864.6 &237.2 &0.05 &59.2 &0.022 &$15D$ &$10D$ &$10D$ &$\Delta x = D/33$ &$\Delta x$ &$\Delta x $\\
        \hline
        864.6 &237.2 &0.05 &59.2 &0.022 &$15D$ &$10D$ &$10D$ &$\Delta x = D/33$ &$\Delta x$ &$\Delta x / 5$\\
        \hline
        864.6 &237.2 &0.05 &59.2 &0.022 &$15D$ &$10D$ &$10D$ &$\Delta x = D/33$ &$\Delta x$ &$\Delta x / 10$\\
        \hline
        \end{tabular}
        \label{table:DropletCollisionMeshSensitivity}
\end{table}
{The results, in terms of slices of the Lagrangian front on the plane of the relative motion, are reported in Figure \ref{fig:DropletCollisionFlow_mesh_sensitivity}.}
{As expected, an increase in the threshold $\delta$ anticipates the first coalescence, which occurs at $t^* \approx 0.0168$ for the case with $\delta \approx \Delta x$. However, such a difference does not significantly affect the solution until $t^* \approx 0.0308$, where a second topology change event takes place for the $\delta = \Delta x$ case only. A breakup is responsible for the formation of a central hole, which connects the external sides of the interface. On the other hand, the cases with $\delta = \Delta x /5, \Delta x /10$ show the formation of two highly-curved regions, where a strong surface tension force locally changes the momentum of the fluid. This interfacial effect is then responsible for the different dynamics that the gas-liquid interfaces exhibit for the remaining part of the simulation. A counter-intuitive effect is shown in Figure \ref{fig:DropletCollisionFlow_mesh_sensitivity} for $t^* = 0.0607$, where the breakup of the right satellite is delayed for the largest threshold case (i.e., $\delta = \Delta x$). This is a consequence of the previous interfacial dynamics, which deviates significantly from the other two cases for $t^* > 0.0308$. The simulations with $\delta = \Delta x /5, \Delta x /10$ produce a very similar behaviour, where no appreciable differences are observed. The differences among these plots and those where we compare against the results of \citet{Finotello2018} (Figure \ref{fig:Finotello2018_fig8d}) occur because, in the latter, we project the front on the plane (in order to obtain consistent visualisations with the experimental images) whilst, in the former, we perform slices that allow to visualise the region inside the interface.}

{It is finally noted that, in the proposed framework, it is sufficient to refine the threshold distance $\delta$ to achieve a solution that matches the experimental results, whilst both the Eulerian and marching cubes grid resolutions are kept constant. On the other hand, traditional VOF/LS methods, which use the same numerical grid for the solution of the Navier-Stokes equations and transport of the interface, would require a finer mesh resolution to achieve a more accurate description of topology changes, resulting in a less efficient approach.}

\subsection{Breakup of a rising bubble}
\label{sec:Breakup of a rising bubble}
In this section we present two cases of three-dimensional rising bubbles that undergo central and peripheral breakup \citep{Tripathi2015, Sharaf2017}. A spherical bubble (with diameter $D$) is initially placed at the bottom of a large domain with dimensions $10D \times 30D \times 10D$ and gravity is applied along the $y$-direction, i.e., $\mathbf{g} = -g \mathbf{\hat{j}}$. The bubble centre coordinates at $t=0$ are $(0, D, 0)$. As usual, the gas-liquid system is characterised by its density $(\rho_\mathrm{r})$ and viscosity $(\mu_\mathrm{r})$ ratios. Two additional non-dimensional numbers are necessary to describe the problem. In this work, we adopt the Galilei number:
\begin{equation}
    \mathrm{Ga}^2 = \frac{g D^3}{\nu_\mathrm{c}^2}
\end{equation}
which describes the ratio between gravitational and viscous forces, and the Bond number:
\begin{equation}
    \mathrm{Bo} = \frac{\rho_\mathrm{c} g D^2}{\sigma}
\end{equation}
which compares gravitational and surface tension effects. \citet{Tripathi2015} present a numerical study on the behaviour of rising bubbles in a gas-liquid system with $\rho_\mathrm{r} = 10^{-3}, \mu_\mathrm{r} = 10^{-2}$ and different sets of $(Ga, Bo)$. The authors identify five different regimes in terms of shapes, rising trajectory and, eventually, breakup. Two different mechanisms that lead to topology changes are observed, namely peripheral and central breakup. In the first regime, the bubble develops a spherical cap followed by an elongated skirt. As the bubble rises, the skirt structure becomes thinner and breaks up releasing downstream satellite bubbles/ligaments. In the central breakup case, the bottom side of the bubble moves upwards faster than the top side and breaks through. A central hole is formed and the bubble deforms into a toroidal shape. Here, we model both breakup mechanisms and a summary of the parameters used in our simulations is shown in Table \ref{table:RisingBubble}.
\begin{table}[ht]
\caption{Non-dimensional governing parameters for droplet-droplet collision.}
\centering
 \begin{tabular}{l c c c c c c c >{\raggedright\arraybackslash}p{2.cm} >{\raggedright\arraybackslash}p{2.cm} c} 
 \hline
 Breakup &$\rho_\mathrm{r}$ &$\mu_\mathrm{r}$ &$\mathrm{Ga}$ &$\mathrm{Bo}$ &$L_x$ &$L_y$ &$L_z$ &Eulerian grid $(\Delta x)$ &Marching cubes grid &$\delta$\\ [0.5ex] 
 \hline
 Peripheral &$10^{-3}$ &$10^{-2}$ &56.6 &339.0 &$10D$ &$30D$ &$10D$ &$\Delta x = D/32$ &$\Delta x$ &$\Delta x$\\
 Central &$10^{-3}$ &$10^{-2}$ &339.4 &400.0 &$10D$ &$30D$ &$10D$ &$\Delta x = D/32$ &$\Delta x$ &$\Delta x / 2$\\
 \hline
 \end{tabular}
 \label{table:RisingBubble}
\end{table}
According to \citet{Tripathi2015}, the selected configurations in terms of Galilei and Bond numbers covers both types of breakup regimes (note that the authors in \citet{Tripathi2015} use the bubble radius instead of the diameter for $\mathrm{Ga}$ and $\mathrm{Bo}$).

\begin{figure}[ht]
    \centering
    \includegraphics[]{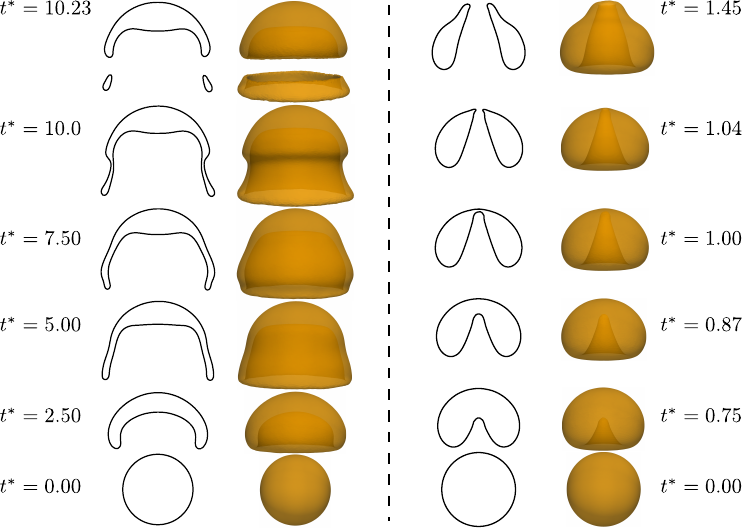}
    \caption{Breakup of a rising bubble with $\mathrm{Ga} = 56.6$, $\mathrm{Bo} = 339$ (left) and $\mathrm{Ga} = 339.4$, $\mathrm{Bo} = 400$ (right). Time is made non-dimensional with $t^* = t \sqrt{g/D}$. The opacity of the three-dimensional visualisations is decreased to ease the visualisation of the inner interface.}
    \label{fig:slices_run1_run2_w3D}
\end{figure}

The results of our simulations are shown in Figure \ref{fig:slices_run1_run2_w3D} for peripheral (left) and central (right) breakup. The figure shows slices of the fronts on the $x-y$ plane {(along with the corresponding three-dimensional visualisations)} at different times (made non-dimensional with the reference time $\sqrt{D/g}$).
Peripheral breakup occurs after the bubble has developed an elongated axi-symmetric skirt structure that fluctuates as the bubble rises. At $t^* = 10.0$, the skirt thickness reaches a minimum located halfway between the spherical cap and the bottom of the bubble. The topology changes algorithm detects breakup and applies marching cubes for front reconstruction. This results into two separated bodies, a main front at the top and a toroidal structure at the bottom, which is unstable and can eventually undergo further breakups. The reconstructed topology of the interface after breakup is consistent with the experimental investigations reported by \citet{Sharaf2017} for the peripheral regime. Central breakup occurs during the first stages of the simulation $(t^* \approx 1.0)$ as the bottom side of the bubble approaches the top side quickly and the distance between the two is smaller than the threshold $\delta$. {This behaviour agrees well with the numerical simulations reported in \citet{Tripathi2015}. However, such a central breakup mechanism, is not observed in the experimental study of \citet{Sharaf2017}. The authors discuss possible reasons behind this discrepancy, such as the initial bubble shape (which is not perfectly spherical in the experiments), different density and viscosity ratios as well as liquid contamination.} 

\subsubsection{A more complex scenario}
\label{sec:A more complex scenario}
\begin{figure}[ht]
        \centering
        \includegraphics[]{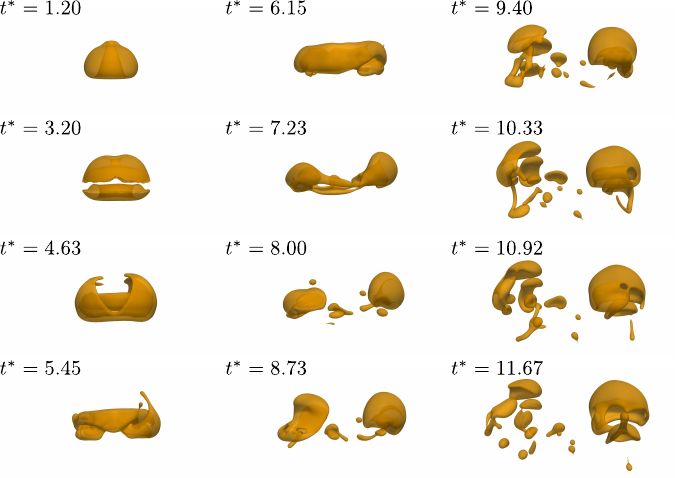}
        \caption{Multiple breakups/merging of a rising bubble following an initial central breakup regime. The opacity of the rendering is decreased to ease the visualisation of the inner interface.}
        \label{fig:ComplexRisingBubble}
\end{figure}

{The two-phase systems analysed in sections \ref{sec:Droplet breakup in simple shear flow}, \ref{sec:Droplet-droplet collision} and \ref{sec:Breakup of a rising bubble} are relatively simple configurations where only a few topology change events occur. Although such scenarios do occur in several applications, (e.g., microfluidics \citep{Salari2020} or surfactant-laden flows \citep{Pico2024}) and are therefore relevant from a practical point of view, we test here the proposed methodology against a more general case where multiple breakup/coalescence events occur and lead to very complex interfacial configurations. The selected test case is the same as the rising bubble with central breakup proposed in section \ref{sec:Breakup of a rising bubble} (see Table \ref{table:RisingBubble}) for the setup, but it is let run for more iterations. The results of this simulation are shown in Figure \ref{fig:ComplexRisingBubble}.}
{A second breakup event occurs at $t^* \approx 3.15$ and resembles a peripheral breakup, where the interface is split into a main cap (on top) and a toroidal structure at the bottom. The shape deviates from an axisymmetrical configuration at $t^* \approx 4.53$, where the bottom structure rises faster than the top one and breaks through it. The following dynamics is characterised by a very complex arrangements of multiple breakups and coalescences that lead to a disperse flow with several satellites. This test case shows the robustness of the proposed framework and its ability to deal with very complex scenarios in two-phase flows with topology changes.}

It is finally reminded that, in the presented validation test cases, the threshold distance $\delta$ is set to a variable fraction of the Eulerian grid $\Delta x$. However, the proposed methodology allows for a seamless implementation of different physics-informed criteria for breakup/coalescence, according to the specific problem under investigation. 
    
\section{Conclusions}
\label{sec:conclusions}
We have presented a numerical method for topology changes (breakup and coalescence of interfaces) in three-dimensional two-phase flows with a front tracking framework. The proposed approach is based on the marching cubes algorithm for triangulation of the iso-surface corresponding to the indicator function value $\mathcal{I} = 0.5$. Breakup and coalescence are obtained by replacing the fluid entities with the corresponding shape as is given by the indicator function field, thus resulting in automatic topology changes in a front capturing way. However, compared to standard VOF or LS methods, the proposed approach offers a better control over the breakup/coalescence of interfaces, as the distance at which these occur is an input of the algorithm that, in principle, can be set to any user-defined criteria and is not necessarily constrained to the resolution of the mesh used to solve the governing equations. To reduce the computational effort of the module for topology changes, marching cubes is executed only for the bodies that potentially undergo topology changes. Fundamental to this is the detection algorithm for breakup/coalescence that uses bounding boxes to perform fast search operations on the triangular fronts. 

The proposed methodology has been tested on several validation cases for topology changes in two-phase disperse flows: droplet in a simple shear flow, droplet-droplet collision and rising bubbles. The breakup of a spherical droplet in a shear flow at low Reynolds number is compared against an experimental work and a good accuracy is found in terms of droplet shapes and size distribution after breakup. The rupture of the interface is also observed in the rising bubble cases, where the two regimes described in the literature (i.e., central and peripheral breakups) are correctly reproduced. Finally, the collision between two equal droplets is used as a validation case for coalescence of two separated interfaces. We have replicated a reflexive separation regime, where the interface breaks up again after the first coalescence and generates three different droplets. Such interfacial flow is characterised by a complex transient dynamics with continuous changes in shapes and orientations. Comparison against experiments shows a very good accuracy in predicting the evolution of the interface. The proposed methodology is second-order accurate for volume conservation, consistently with the numerical schemes adopted to spread the indicator function from the Lagrangian front to the Eulerian mesh. The marching cubes grid resolution as well as breakup/coalescence threshold $\delta$ have a central role in the proposed framework and an accurate choice of these parameters is a crucial aspect. This emphasises the importance of coupling numerical frameworks for topology changes with physics-informed criteria to obtain accurate results.  

In the context of front tracking simulations for two-phase flows with topology changes, the proposed methodology offers a robust approach to model very complex changes in the shape of interfaces. In this way, we overcome the limitations of grid-free methods that operate only on the front elements that are close to the area affected by topology-changes and are not robust when large portions of the front undergo breakup/coalescence. Compared to other grid-based approaches (e.g., LCRM, LFRM), our methodology offers the advantage of restricting front reconstruction operations only for those bodies that undergo topology-changes, reducing the computational cost and, at the same time, mitigating the error introduced by front reconstruction operations. The proposed methodology provides a robust framework for front tracking simulations of complex two-phase flows with topology changes and paves the way towards a better, physics-based, modelling of interfacial flows undergoing breakup and coalescence. 

\section*{Data Availability Statement}
The data that support the findings of this study are reproducible
and data is openly available in the repository with DOI
10.5281/zenodo.15075654, available at \url{https://doi.org/10.5281/zenodo.15075654}

\section*{Acknowledgements}
\noindent This project has received funding from the Deutsche Forschungsgemeinschaft (DFG, German Research Foundation), grant numbers 420239128 and 458610925.

\bibliographystyle{model1-num-names}


\end{document}